\documentclass[prd,aps,preprint,tightenlines,floats,epsfig,superscriptaddress]{revtex4}
\usepackage{epsfig}
\usepackage{amsfonts}
\usepackage{amsmath}
\usepackage{graphicx}
\begin{document}

\preprint{\vbox{\hbox{UMD-DOE-40762-465}}}

\title{P-odd and CP-odd Four-Quark Contributions to Neutron EDM}

\author{Haipeng An}
 \affiliation{Maryland Center for Fundamental Physics and Department of Physics, University of
Maryland, College Park, Maryland 20742, USA }
\author{Xiangdong Ji}
 \affiliation{Maryland Center for Fundamental Physics and Department of Physics, University of
Maryland, College Park, Maryland 20742, USA }
 \affiliation{Institute of Particle Physics and Cosmology,
 Department of Physics, Shanghai Jiao Tong University, Shanghai 200240, China}
 \affiliation{Center
for High-Energy Physics and Institute of Theoretical Physics, Peking
University, Beijing 100871, China}
\author{Fanrong Xu}
 \affiliation{Institute of Theoretical Physics, Chinese Academy of Sciences, Beijing 100190, China}
\date{\today}
\vspace{0.8in}

\begin{abstract}

In a class of beyond-standard-model theories, CP-odd observables,
such as the neutron electric dipole moment, receive significant
contributions from flavor-neutral P-odd and CP-odd four-quark
operators. However, considerable uncertainties exist in the hadronic
matrix elements of these operators strongly affecting the
experimental constraints on CP-violating parameters in the theories.
Here we study their hadronic matrix elements in combined chiral
perturbation theory and nucleon models. We first classify the
operators in chiral representations and present the leading-order
QCD evolutions. We then match the four-quark operators to the
corresponding ones in chiral hadronic theory, finding symmetry
relations among the matrix elements. Although this makes lattice QCD
calculations feasible, we choose to estimate the non-perturbative
matching coefficients in simple quark models. We finally compare the
results for the neutron electric dipole moment and P-odd and CP-odd
pion-nucleon couplings with the previous studies using naive
factorization and QCD sum rules. Our study shall provide valuable
insights on the present hadronic physics uncertainties in these
observables.

\end{abstract}

\maketitle

\section{Introduction}

Neutron electric dipole moment (nEDM) has attracted considerable attention over more
than half a century. For an elementary particle to have non-vanishing intrinsic EDM, simple analysis
shows that parity-violating as well as time-reversal-violating interactions must be present.
[T-violation is equivalent to CP-violation (combined charge-conjugation and parity) in local quantum field theory.]
However, in the standard model (SM) of particle physics, such interactions arise only
from flavor-changing Cabbibo-Kobayashi-Moskawa (CKM) matrix elements, which are strongly
suppressed phenomenologically, yielding a very small neutron EDM of order $10^{-31}$ ecm.
Therefore, an experimental observation of a large-size neutron EDM is an unambiguous
signal for new physics, widely expected to exit somewhere
between the electroweak symmetry breaking and TeV scales.

There has been much speculation in the literature on the nature of
new physics. The leading theory is supersymmetry (SUSY) introduced
to stabilize the electroweak scale against the fundamental Planck
scale (for a good review, see Ref.~\cite{Martin:1997ns}). Various
extra dimensional models have been popular in recent years as well
(for good reviews see Ref. \cite{Sundrum:2005jf}). There are other
less dramatic possibilities including
technicolor~\cite{Lane:2002wv}, left-right
symmetry~\cite{Mohapatra:1979ia}, minimal extensions of the
SM~\cite{Lee:1973iz}, little Higgs models~\cite{ArkaniHamed:2001nc},
etc. In each of these models, there is new CP violation physics
giving rise to a significant neutron EDM. To test CP-violation
mechanisms, it is important to get accurate predictions of
CP-violating observables in these models. In particular, one has to
deal with the non-perturbative quantum chromodynamics (QCD) physics
present in the structure of the neutron.

An efficient way to calculate the neutron EDM is to use the methodology of
effective field theories (EFT). In this approach, one generates P-odd and CP-odd
quark and gluon operators after integrating out the heavy particles (including
heavy quarks, gauge bosons and new particles) and run these operators
to a scale around 1 GeV where non-perturbative QCD physics becomes important.
The effective degrees of freedom involves the light quarks (up, down and strange)
and gluons. The CP-odd part of the lagrangian is generally written as
a sum of CP-odd operators of different mechanical dimensions,
\begin{equation}
    {\cal L}_{\rm CP-odd} = \sum_{d=3}^{\infty} \sum_i C_{di}(\mu)\hat O_{di}(\mu)  \ ,
\end{equation}
where $d=3, 4, 5,$ etc, is the mechanical dimension of the
operators, $\mu$ is the renormalization scale (taken as $4\pi F_\pi$
in this paper) and $i$ sums over operators of the same dimension.
The dim-3 operator is the usual CP-odd quark mass term $\bar q
i\gamma_5 q$, which can be rotated away through chiral rotations
apart from the $U_A(1)$ anomaly. The dim-4 operator is the usual
$\theta$ term $G\tilde G$.  Dim-5 operators include quark electric
and chromoelectric dipole operators. Dim-6 operators contain various
four-quark operators and Weinberg three-gluon operator. The matrix
elements of dim-4 and, to less extent, dim-5 operators have been
studied extensively in the
literature~\cite{Baluni:1978rf,Crewther:1979pi,Pich:1991fq,Pospelov:1999ha,
Pospelov:1999mv,He:1989xj,Beall:1981zq,Abel:2004te,Pospelov:2000bw,Hockings:2005cn},
and the uncertainty of the estimates are typically at the level of
factor 2. The contributions of these operators have also been
studied extensively in the context of various new physics models
(see Refs.~\cite{Erler:2004cx,Pospelov:2005pr} for good reviews).

However, the matrix elements of dim-6 operators have been a
challenge to estimate. In some beyond-SM theories such as the
left-right symmetric model, dim-6 four-quark operators dominate the
contributions to nEDM. In the literature, the only serious approach
that has been proposed to calculate their matrix elements is the
naive factorization method: breaking the four-quark matrix elements
into the product of two-quark matrix elements between the nucleon
states and between pion and
vacuum~\cite{Khatsimovsky:1987fr,Valencia:1989cm,He:1992db}. While
the factorization involving mesons can be and has been tested using
lattice QCD~\cite{Babich:2006bh} and the results may be trustable to
within a factor of 2, the same is not known for matrix elements
involving the nucleon states. The goal of this paper is to develop a
chiral perturbation method combined with simple quark models to
estimate the four-quark contribution to the nEDM with hopefully an
improved accuracy.

The approach we are going to take is the standard chiral
perturbation theory ($\chi$PT) (see, for example,
Ref.~\cite{Scherer:2002tk,Bernard:1995dp}) which has been used to
calculate the contribution of $\theta$-term to
nEDM~\cite{Pich:1991fq}. One of the successes of the chiral approach
can be illustrated by the polarizabilities of the nucleon. The
electric polarizabilities of the proton and neutron have been
extracted from experimental data, $\alpha_p^{\rm exp}  =(10.4\pm
0.6) \times 10^{-4}{\rm fm}^3$, $\alpha_n^{\rm exp} =(12.3\pm 1.3)
\times 10^{-4}{\rm fm}^3$. The leading contribution in $\chi$PT
comes from the pion-nucleon intermediate states,
\begin{equation}
   \alpha_p = \alpha_n = \frac{5 \alpha_{\rm em} g_A^2}{96\pi F_\pi^2 m_\pi} \approx
   11\times10^{-4} {\rm fm}^3 \ ,
\end{equation}
which diverges linearly as $m_\pi \rightarrow 0$ and agrees well
with the experimental data. One would expect then a similar pion
dominance in the neutron EDM because the latter also involves the
intermediate electric dipole excitations. Indeed a pioneering
calculation by Crewther et al. found that the dominant contribution
from the charged-pion chiral-loop diverges logarithmically as
$m_\pi$ goes to zero, and is proportional to the CP-odd
pion-nucleon-nucleon coupling $\bar g_{\pi
NN}$~\cite{Crewther:1979pi}. In this paper, we take this
contribution as dominating and consider the four-quark operator
contribution to $\bar g_{\pi NN}$. Of course, there are
chiral-regular contributions to the nEDM which are of the same order
in chiral power counting and numerically competitive or even
dominating in the real world~\cite{Pospelov:2005pr}. We will
consider these contributions as well, although the model-dependence
becomes unavoidable.

In the chiral approach, one first writes down the CP-odd and even
lagrangian in terms of meson and nucleon fields,
\begin{eqnarray}
  {\cal L} &=&  {\cal L}_{\rm Goldstone-boson~CP-odd~term} +  {\cal L}_{\rm nucleon~CP-odd~mass~term} \nonumber \\
  && + {\cal L}_{\rm EDM~term} + {\cal L}_{\rm CP-odd~\pi-N~coupling} + ({\rm CP-even~terms})
\end{eqnarray}
where the Goldstone boson CP-odd lagrangian will generate terms
annihilating $\pi^0$ and $\eta$ in the vacuum, or in other words,
will produce meson condensates. The condensates will turn some of
the CP-even terms (as we shall see, those proportional to quark
masses) in the chiral lagrangian into CP-odd contributions. This
will generate an additional CP-odd nucleon-mass term, neutron EDM
term and CP-odd pion-nucleon coupling. Once this is done, one can
rotate away the CP-odd nucleon mass term, generating further
contributions to the neutron EDM terms and the CP-odd pion-nucleon
coupling.

The presentation of the paper is organized as follows: In Sec. II,
we classify all flavor-neutral P-odd and CP-odd four-quark operators
in chiral representations. We also present the leading-order QCD
scale evolution of these operators. In Sec. III, we match these
operators to the corresponding Goldstone boson operators, baryon
operators, and EDM operators in $\chi$PT. We also discuss in the
case of Peccei-Quinn symmetry the size of the induced $\theta$ term
in the presence of these four-quark operators. In Sec. IV, we
calculate their contributions to the P-odd and CP-odd nucleon-pion
vertices and the CP-odd nucleon mass using factorization in the case
of meson matrix elements and simple quark models for the nucleon
ones. In Sec. V, we study the four-quark contribution to the neutron
EDM in the chiral approach supplemented with factorization and quark
model estimates of counter terms, and the results are compared with
other calculations in the literature. The comparison and analysis
show that the hadronic physics uncertainties here can be quantified
to within a factor of two for operators generating unsuppressed
meson condensate contributions. We conclude the paper in Sec.VI.

To make the convention clear, the EDM operator for spin-1/2
particles is defined as
\begin{equation}
{\cal L} = -\frac{1}{2}d^E \psi \sigma_{\mu\nu} i\gamma_5 \psi
F^{\mu\nu},
\end{equation}
where $\sigma_{\mu\nu} = \frac{i}{2}[\gamma_\mu,\gamma_\nu]$ and $\gamma_5 = i\gamma^0\gamma^1\gamma^2\gamma^3$ in
the standard Dirac representation.

\section{P-odd and CP-odd Four-quark operators: Classification, running and mixing}

We consider three light quark flavors: up, down and strange.
Flavor-neutral P-odd and CP-odd four-quark operators can be divided
into two groups \cite{Khatsimovsky:1987fr}: The first group includes
18 operators made of two different flavors:
\begin{eqnarray}\label{4quark}
O_{11} &=& (\bar q i\gamma_5 q)(\bar q' q') \ , \nonumber\\
O_{12} &=& (\bar q q)(\bar q' i\gamma_5 q') \ , \nonumber\\
O_{21} &=& (\bar q i\gamma_5 t^a q)(\bar q' t^a q')\ , \nonumber\\
O_{22} &=& (\bar q t^a q)(\bar q' i\gamma_5 t^a q')\ , \nonumber\\
O_{3} &=& (\bar q i\gamma_5 \sigma^{\mu\nu} q)(\bar q'
\sigma_{\mu\nu} q')  \ , \nonumber\\
O_{4} &=& (\bar q i\gamma_5 \sigma^{\mu\nu} t^a q)(\bar q'
\sigma_{\mu\nu} t^a q') \ ,
\end{eqnarray}
where $q, q'=u,d,s$ and $q< q'$. The second group includes 6 operators
made of a single quark flavor
\begin{eqnarray}\label{4quark2}
O'_1 &=& (\bar q i\gamma_5 q)(\bar q q) \ , \nonumber\\
O'_2 &=& (\bar q i\gamma_5 t^a q)(\bar q t^a q) \ ,
\end{eqnarray}
where $t^a$ are generators of the $SU(3)$ color group. All other
flavor-neutral P-odd and CP-odd four-quark operators can be related
to the above through Fierz transformation
\cite{Khatsimovsky:1987fr}.

To match the above quark operators into the hadronic ones in $\chi$PT,
we have to classify the former into irreducible representations of the chiral
group $SU(3)_L\times SU(3)_R$. Take the operator $\bar u i\gamma_5
u\bar d d$ as an example, which can be decomposed as
\begin{equation}\label{3}
\bar u i\gamma_5 u\bar d d=-i \bar u_R u_L\bar d_R d_L +i \bar u_L
u_R\bar d_R d_L \;+\;{\rm h.c.} \ ,
\end{equation}
where $q_{L,R} = P_{L,R} q$ with $P_{L,R} = (1\mp \gamma_5)/2$.
The first term can be further decomposed as
\begin{eqnarray}\label{4}
-i\bar u_R u_L \bar d_R d_L&=&-\frac{i}{2}(\bar u_R u_L \bar d_R d_L
+\bar d_R u_L \bar u_R d_L)\nonumber\\
&&-\frac{i}{2}(\bar u_R u_L \bar d_R d_L -\bar d_R u_L \bar u_R
d_L)\nonumber\\
&=&-\frac{i}{4}S^{ij}_{kl}\bar q_{Ri}q_{L}^k \bar q_{Rj}
q_L^l-\frac{i}{4}\epsilon^{jkl}_{imn}A^i_j\bar q_{Rk}q_L^m\bar
q_{Rl}q_L^n\ ,
\end{eqnarray}
where $\epsilon^{jkl}_{imn}\equiv\epsilon^{jkl}\epsilon^{imn}$, and
\begin{equation}\label{5}
A=\left(\begin{array}{ccc}0&0&0\\0&0&0\\0&0&1\\
\end{array}\right),
\end{equation}
and
\begin{equation}\label{6}
S^{12}_{12}=S^{21}_{12}=S^{12}_{21}=S^{21}_{21} = 1,
\end{equation}
with other elements vanishing. The second term of Eq. (\ref{3}) can
be written as
\begin{equation}\label{7}
i\bar u_L u_R \bar d_R d_L=iH^j_{1i}H^l_{2k}\bar
q_{Lj}q_{R}^k\bar q_{Rl}q_L^i\;,
\end{equation}
where
\begin{equation}\label{8}
H_1=\left(\begin{array}{ccc}0&1&0\\0&0&0\\0&0&0\\
\end{array}\right)\;,\;\;\;H_2=\left(\begin{array}{ccc}0&0&0\\1&0&0\\0&0&0\\
\end{array}\right)\;.
\end{equation}
In this way the operator $\bar u i\gamma_5 u\bar d d$ is decomposed
into $(\bar 3,3)$, $(6,\bar6)$ and $(8,8)$ representations of
$SU(3)_L\times SU(3)_R$, and $A$, $S$, $H_1$, and $H_2$ can be regarded
as spurion fields in the sense that if they transform as $(3,\bar
3)$, $(\bar 6,6)$ and $(8,8)$  under chiral transformation, the corresponding
terms in Eqs. (\ref{4}) and (\ref{7}) become invariant. These spurion fields will
be used in $\chi PT$ to construct the effective operators corresponding to the same
four-quark operators. All spurion fields for four-quark operators with different
Dirac and color structures are shown in Table \ref{spurions1}. [It is easy to see that there is
no $(1,1)$ operator because any such operator must be expressible in terms of
products of chiral-even quark currents, which cannot yield CP-odd contributions.]

\begin{table}[h]
\begin{tabular}{|c|c|c|c|}
\hline
~&~$(\bar 3,3)$~&~$(6,\bar 6)$~&~$(8,8)$ \\
\hline
$q=u,\;q'=d$~&~$A=\left(\begin{array}{ccc}0&0&0\\0&0&0\\0&0&1\\ \end{array}\right)$~&~$\begin{array}{c}S^{12}_{12}=S^{12}_{21}=S^{21}_{12}=S^{21}_{21}=1\\ {\rm other\;\; components\;\; are\;\; zero}\\ \end{array}$~&~$H_1=\left(\begin{array}{ccc}0&1&0\\0&0&0\\0&0&0\\ \end{array}\right),\;\;H_2=\left(\begin{array}{ccc}0&0&0\\1&0&0\\0&0&0\\ \end{array}\right)$  \\
\hline
$q=d,\;q'=u$~&~$A=\left(\begin{array}{ccc}0&0&0\\0&0&0\\0&0&1\\ \end{array}\right)$~&~$\begin{array}{c}S^{12}_{12}=S^{12}_{21}=S^{21}_{12}=S^{21}_{21}=1\\ {\rm other\;\; components\;\; are\;\; zero}\\ \end{array}$~&~$H_1=\left(\begin{array}{ccc}0&0&0\\1&0&0\\0&0&0\\ \end{array}\right),\;\;H_2=\left(\begin{array}{ccc}0&1&0\\0&0&0\\0&0&0\\ \end{array}\right)$  \\
\hline
$q=u,\;q'=s$~&~$A=\left(\begin{array}{ccc}0&0&0\\0&1&0\\0&0&0\\ \end{array}\right)$~&~$\begin{array}{c}S^{13}_{13}=S^{13}_{31}=S^{31}_{13}=S^{31}_{31}=1\\ {\rm other\;\; components\;\; are\;\; zero}\\ \end{array}$~&~$H_1=\left(\begin{array}{ccc}0&0&1\\0&0&0\\0&0&0\\ \end{array}\right),\;\;H_2=\left(\begin{array}{ccc}0&0&0\\0&0&0\\1&0&0\\ \end{array}\right)$  \\
\hline
$q=s,\;q'=u$~&~$A=\left(\begin{array}{ccc}0&0&0\\0&1&0\\0&0&0\\ \end{array}\right)$~&~$\begin{array}{c}S^{13}_{13}=S^{13}_{31}=S^{31}_{13}=S^{31}_{31}=1\\ {\rm other\;\; components\;\; are\;\; zero}\\ \end{array}$~&~$H_1=\left(\begin{array}{ccc}0&0&0\\0&0&0\\1&0&0\\ \end{array}\right),\;\;H_2=\left(\begin{array}{ccc}0&0&1\\0&0&0\\0&0&0\\ \end{array}\right)$  \\
\hline
$q=d,\;q'=s$~&~$A=\left(\begin{array}{ccc}1&0&0\\0&0&0\\0&0&0\\ \end{array}\right)$~&~$\begin{array}{c}S^{23}_{23}=S^{23}_{32}=S^{32}_{23}=S^{32}_{32}=1\\ {\rm other\;\; components\;\; are\;\; zero}\\ \end{array}$~&~$H_1=\left(\begin{array}{ccc}0&0&0\\0&0&1\\0&0&0\\ \end{array}\right),\;\;H_2=\left(\begin{array}{ccc}0&0&0\\0&0&0\\0&1&0\\ \end{array}\right)$  \\
\hline
$q=s,\;q'=d$~&~$A=\left(\begin{array}{ccc}1&0&0\\0&0&0\\0&0&0\\ \end{array}\right)$~&~$\begin{array}{c}S^{23}_{23}=S^{23}_{32}=S^{32}_{23}=S^{32}_{32}=1\\ {\rm other\;\; components\;\; are\;\; zero}\\ \end{array}$~&~$H_1=\left(\begin{array}{ccc}0&0&0\\0&0&0\\0&1&0\\ \end{array}\right),\;\;H_2=\left(\begin{array}{ccc}0&0&0\\0&0&1\\0&0&0\\ \end{array}\right)$  \\
\hline
$q=u,\;q'=u$~&~$A=0$~&~$\begin{array}{c}S^{11}_{11}=4\\ {\rm other\;\; components\;\; are\;\; zero}\\ \end{array}$~&~$H_1=0,\;\;H_2=0$  \\
\hline
$q=d,\;q'=d$~&~$A=0$~&~$\begin{array}{c}S^{22}_{22}=4\\ {\rm other\;\; components\;\; are\;\; zero}\\ \end{array}$~&~$H_1=0,\;\;H_2=0$  \\
\hline
$q=s,\;q'=s$~&~$A=0$~&~$\begin{array}{c}S^{33}_{33}=4\\ {\rm other\;\; components\;\; are\;\; zero}\\ \end{array}$~&~$H_1=0,\;\;H_2=0$  \\
\hline
\end{tabular}
\caption{Spurions for CP-odd 4-quark operators. The first six
together with three tensor structures yield 18 operators in Eq. (4)
and the last three with two tensor structures yield six operators in
Eq. (5).}\label{spurions1}
\end{table}

The four-quark operators usually emerge at a high energy scale where
some heavy particles have been integrated out. To match them to
hadronic operators in effective theories, one must run them down to
a low energy scale where non-perturbative physics becomes important.
We can choose this to be 1 GeV or the lattice cut-off $1/a$, where
$a$ is the lattice spacing. In this work, we take $\mu=4\pi F_\pi$,
with $F_\pi = 93$ MeV. These operators mix with each other when the
energy scale changes. Although many of the mixings have been
calculated in the literature before [see Ref. \cite{Miller:1982ij},
for example], we recalculate them and present the complete result
here for easy reference:
\begin{eqnarray}\label{running}
\mu^2\frac{d}{d\mu^2}\left(\begin{array}{c}O_{11}\\O_{12}\\O_{21}\\O_{22}\\O_{3}\\O_{4}\\ \end{array}\right)&=&\frac{\alpha_S(\mu)}{4\pi}\left(\begin{array}{cccccc}8&0&0&0&0&1\\0&8&0&0&0&1\\0&0&-1&0&\frac{2}{9}&\frac{5}{12}\\
0&0&0&-1&\frac{2}{9}&\frac{5}{12}\\0&0&24&24&-\frac{8}{3}&0\\
\frac{16}{3}&\frac{16}{3}&10&10&0&\frac{19}{3}\\ \end{array}\right)\left(\begin{array}{c}O_{11}\\O_{12}\\O_{21}\\O_{22}\\O_{3}\\O_{4}\\
\end{array}\right) \ , \\ \label{running1}
\mu^2\frac{d}{d\mu^2}\left(\begin{array}{c}O'_{1}\\O'_{2}\\
\end{array}\right)&=&\frac{\alpha_S(\mu)}{4\pi}\left(\begin{array}{cc}\frac{40}{9}&-\frac{4}{3}\\-\frac{80}{27}&-\frac{46}{9}\\
\end{array}\right)\left(\begin{array}{c}O'_{1}\\O'_{2}\\
\end{array}\right) \ .
\end{eqnarray}
Clearly operators with different quark flavor structures do not mix.
Since $SU(3)_L\times SU(3)_R$ symmetry is broken only by quark masses,
four-quark operators belonging to different chiral irreducible
representations do not mix either. Therefore, we can further simplify Eq. (12),
\begin{eqnarray}
\mu^2\frac{d}{d\mu^2}\left(\begin{array}{c}O_1^{(3,6)}\\O_2^{(3,6)}\\O_3^{(3,6)}\\O_4^{(3,6)}\\
\end{array}\right)&=&\frac{\alpha_S(\mu)}{4\pi}\left(\begin{array}{cccc}8&0&0&1\\0&-1&\frac{2}{9}&\frac{5}{12}\\0&48&-\frac{8}{3}&0\\ \frac{32}{3}&20&0&\frac{19}{3}\\
\end{array}\right)\left(\begin{array}{c}O_1^{(3,6)}\\O_2^{(3,6)}\\O_3^{(3,6)}\\O_4^{(3,6)}\\
\end{array}\right)\\ \label{12}
\mu^2\frac{d}{d\mu^2}\left(\begin{array}{c}O_1^{(8)}\\O_2^{(8)}\\
\end{array}\right)&=&\frac{\alpha_S(\mu)}{4\pi}\left(\begin{array}{cc}8&0\\0&-1\\
\end{array}\right)\left(\begin{array}{c}O_1^{(8)}\\O_2^{(8)}\\ \end{array}\right)
\end{eqnarray}
where $O_i^{(3,6,8)}$ means the projections of the operator $O_i$ on
the representations $(\bar3,3)$, $(6,\bar6)$ and $(8,8)$,
respectively. It is easy to see that the $(\bar3,3)$ and $(6,\bar6)$
projections of $O_{i1}$ and $O_{i2}$ are the same with $i=1,2$,
whereas their $(8,8)$ projections differ only by the sign. The tensor operators
do not have $(8,8)$ components and therefore do not participate
in Eq. (\ref{12}).

The four-quark operators may also mix with P-odd and CP-odd operators
with dimension less or equal to 6. For mixing with lower-dimensional
operators, either quark masses or power divergences will appear.
The only other dimension-6 operator is the Weinberg operator \cite{Weinberg:1989dx}
\begin{equation}
O_W=-\frac{1}{6}f^{abc}\epsilon^{\mu\nu\alpha\beta}G^a_{\mu\rho}G^{b\rho}_{\nu}G^c_{\alpha\beta},
\end{equation}
which is a singlet under chiral transformation. Since the
four-quark operators contain no singlet component, the mixing between
them and $O_W$ vanishes. The evolution of the
Weinberg operator can be found in Ref.~\cite{Braaten:1990gq}.

The P-odd and CP-odd dimension-5 operators are the quark electric
dipole moment operators (QEDM) and quark chromo-electric dipole
moment operators (QCDM). In principle, they belong to $(\bar 3,3)$
of the chiral group. However, they can mix logarithmically with
four-quark operators multiplied by the quark mass which transforms
also like $(\bar 3,3)$~\cite{Frere:1991jt,Shifman:1976de}.

Finally, the four-quark operators can have mixing with
$m\bar qi\gamma_5 q$ with quadratically divergent coefficients.
Usually, one defines the four-quark operators with quadratic
divergences subtracted, as is natural in dimensional regularization
where all quadratically divergent integrals vanish by definition.
Equivalently, this can be achieved, for example,
by demanding the CP-odd four-quark operators have vanishing contribution
between QCD vacuum and CP-odd meson states in perturbation theory.
However, as we shall see in the following section, they can have
non-perturbative contributions. The exact physical implication of this
non-perturbative contribution will be discussed in Sec. V.E.

\section{Matching to Operators in Chiral Perturbation Theory}

Generically, any P-odd, CP-odd quark-gluon operator contributes to
all P-odd, CP-odd hadronic operators in $\chi$PT; the latter are
constructed in terms of Goldstone-boson (pion, kaon, eta) fields and
baryon fields. Here we consider just the contributions to the
Goldstone-boson CP-odd interactions, nucleon CP-odd mass term,
$\pi$-$N$ CP-odd coupling, as well as the neutron EDM term,
\begin{eqnarray}
{\cal L} &=& {\cal L}_{\rm Goldstone-boson~CP-odd~term} + {\cal L}_{\rm nucleon~CP-odd~mass~term}  \nonumber \\
  && + {\cal L}_{\rm CP-odd~\pi-N~coupling} + {\cal L}_{\rm EDM~term} \ .
\end{eqnarray}
Following the standard practice in the literature, we imbed the Goldstone-boson fields in
the unitary matrix $U=\exp[2i\Sigma/F_\pi]$ with
\begin{equation}\label{meson}
\Sigma=\left(\begin{array}{ccc}\frac{1}{2}\pi^0+\frac{1}{2\sqrt{3}}\eta&\frac{1}{\sqrt{2}}\pi^+&\frac{1}{\sqrt{2}}K^+\\
\frac{1}{\sqrt{2}}\pi^-&-\frac{1}{2}\pi^0+\frac{1}{2\sqrt{3}}\eta&\frac{1}{\sqrt{2}}K^0\\
\frac{1}{\sqrt{2}}K^-&\frac{1}{\sqrt{2}}\bar
K^0&-\frac{1}{\sqrt{3}}\eta\end{array}\right) \ ,
\end{equation}
where $F_\pi$ is the pion decay constant. Under chiral rotations,
$U$ transforms like $U\rightarrow LUR^\dagger$, where $L$ and $R$
are SU(3) matrices belonging to $SU(3)_L$ and $SU(3)_R$ groups,
respectively.

To include the baryon octet, we introduce
\begin{equation}
B=\left(\begin{array}{ccc}\frac{1}{\sqrt{2}}\Sigma^0+\frac{1}{\sqrt{6}}\Lambda&\Sigma^+&p\\
\Sigma^-&-\frac{1}{\sqrt{2}}\Sigma^0+\frac{1}{\sqrt{6}}\Lambda&n\\
\Xi^-&\Xi^0&-\frac{2}{\sqrt{6}}\Lambda\\ \end{array}\right) \ .
\end{equation}
Again following the literature, we assume $B$ transforms nonlinearly
under chiral transformation,
\begin{equation}
    B \rightarrow  KBK^\dagger
\end{equation}
where $K$ is a unitary matrix defined according to the transformation of $\xi=U^{1/2}$.
\begin{equation}
     \xi \rightarrow  L\xi K^\dagger, ~~~~  \xi \rightarrow K\xi R^\dagger
\end{equation}
It is clear that $K$ is a nonlinear function of the Goldstone-boson fields.

The quark-mass term breaks chiral symmetry and plays an important role
in chiral expansion. To exhibit its physical effect, the usual practice is
to introduce the spurion field $\chi$, transforming as
\begin{equation}
          \chi \rightarrow  L\chi R^\dagger \ .
\end{equation}
However, to combine $\chi$ with the baryon field $B$, we introduce $\chi_\pm$
\begin{equation}
       \chi_\pm =   \xi^\dagger \chi \xi^\dagger \pm \xi \chi^\dagger \xi \ ,
\end{equation}
which transform nonlinearly as $\chi_\pm \rightarrow K\chi_\pm K^\dagger$.

In the leading order, the chiral lagrangian for meson fields is
\begin{equation}\label{mesonlagrangian}
{\cal L}=\frac{1}{4}F_\pi^2{\rm Tr}[\partial_\mu U^\dagger
\partial^\mu U] + \frac{1}{2}F_\pi^2 B{\rm Tr}[M^\dagger U + U^\dagger
M]\;,
\end{equation}
where $M=diag\{m_u,m_d,m_s\}$ is the mass matrix of light quarks.
The leading-order chiral lagrangian for the baryon field
is~\cite{Bernard:1995dp}
\begin{equation}
  {\cal L} = {\rm Tr}\left\{\bar B i\gamma^\mu D_\mu B - m_0 \bar B B + \frac{1}{2} D \bar B\gamma^\mu\gamma_5{u_\mu,B} + \frac{1}{2}F\bar
  B\gamma_\mu\gamma_5[u_\mu,B]\right\}\;,
\end{equation}
where $u_\mu = i(\xi^\dagger\partial_\mu \xi - \xi\partial_\mu
\xi^\dagger)$ is an axial vector current, $D_\mu B = \partial_\mu B + [\Gamma_\mu, B]$ and
$\Gamma_\mu = \{\xi^\dagger\partial_\mu\xi+\xi\partial_\mu\xi^\dagger\}/2$ is a vector current.

\subsection{Matching to CP-Odd Goldstone-Boson Operators}

Once there is a CP-odd term in the QCD lagrangian, it induces CP-odd
terms in the effective Goldstone-boson lagrangian. These terms can
annihilate odd-number (particularly, one) Goldstone bosons into the vacuum.
Because of this CP-odd meson condensate, the original CP-even terms
can now contribute to the CP-odd effects. Due to chiral symmetry,
a meson condensate can generate physical effects only when the CP-even terms
explicitly break the symmetry.

As discussed in the last section, P-odd and CP-odd four-quark
operators can be decomposed into chiral $(\bar 3,3)$, $(6,\bar6)$,
$(8,8)$ and their hermitian conjugate representations. They in turn
can be matched to the corresponding chiral operators in the same
representations. The leading ones without derivatives are unique and
are shown in Table \ref{mesonoperator}.

\begin{table}[h]

\begin{tabular}{lccc}
\hline  Rep. ~&~ $(\bar3,3)$~&~$(6,\bar6)$~&~$(8,8)$\\
 Operator ~&~ $O^m_3=i{\rm Tr}[A U^\dagger]$ ~&~ $O^m_6=i S^{ij}_{kl}
U^k_i U^l_j$ ~&~ $O^m_8=i{\rm Tr}[H_1 U H_2 U^\dagger] $\\
\hline
\end{tabular}
\caption{Leading meson operators in individual irreducible chiral
representations where $A$, $S$, $H_1$ and $H_2$ are spurion fields
in Table \ref{spurions1}. The appearance of $i$ in front of each
operator indicates that these operators generate P-odd and CP-odd
vertices in the meson lagrangian; their Wilson coefficients in the
lagrangian are defined to be real.}\label{mesonoperator}
\end{table}

We illustrate the matching process using $O_{11}^{ud}=\bar
ui\gamma_5 u\bar d d$ as an example. As discussed in the last
section, this quark operator can be decomposed into irreducible
representations of the chiral group using the spurion fields
\begin{equation}
O_{11}^{ud}\;=\;O_{11}^{ud,(\bar 3,3)}+O_{11}^{ud,(6,\bar
6)}+O_{11}^{ud,(8,8)} \;+\;{\rm h.c.}
\end{equation}
Then, we can match each of the operators to the corresponding one in
the meson sector through the non-perturbative Wilson coefficients
$C's$
\begin{equation}
O_{11}^{ud,(\bar 3,3)}\sim C^{(\bar3,3)} O_3\ ,\;\;
O_{11}^{ud,(6,\bar 6)}\sim C^{(6,\bar6)} O_6\ ,\;\;
O_{11}^{ud,(8,8)}\sim C^{(8,8)} O_8\ .\;\;
\end{equation}
The Wilson coefficients can be obtained by matching the simplest matrix
elements: $\langle 0|O|\pi^0\rangle$ and $\langle
0|O|\eta\rangle$, which can be calculated using non-perturbative
methods such as lattice QCD.

In this paper, we use factorization approximation to estimate these
non-perturbative matrix elements. Lattice QCD calculations demonstrate that
the matrix elements of four-quark operators can be factorized typically to within
a factor of 2. Again take the operator $O_{11}^{ud}$ as an example,
which can annihilate $\pi^0$ and $\eta$ to the vacuum. [In principle, it also
annihilates $\eta'$, but this contribution is suppressed by the
mass of $\eta'$.] The annihilation amplitude can be estimated
using vacuum saturation,
\begin{equation}
\langle0| \bar ui\gamma_5u\bar d d|\pi^0\rangle\approx\langle0|\bar
dd|0\rangle\langle0|\bar ui\gamma_5u|\pi^0\rangle.
\end{equation}
Using chiral symmetry, one can get $ \langle0|\bar
ui\gamma_5u|\pi^0\rangle =\frac{1}{F_\pi}\langle0|\bar uu|0\rangle
\equiv - F_\pi B_0 $, and $ \langle0|\bar ui\gamma_5u|\eta\rangle =
- F_\pi B_0/\sqrt{3} $. (This is consistent with the definition of
the chiral rotation of $U$ defined below Eq. (\ref{meson}).)
Therefore, a term $C_4O_{11}^{ud}$ in the QCD lagrangian can be
matched to the linear terms in $\pi^0$ and $\eta$ in the chiral
lagrangian
\begin{equation}\label{linear}
{\cal L} =C_4B_0^2F_\pi^3\pi^0+\frac{1}{\sqrt{3}}C_4B_0^2F_\pi^3\eta
+ ... \ ,
\end{equation}
where $...$ represents higher-power meson fields. Then the leading terms in the potential of $\pi^0$ and $\eta$
can be written as
\begin{eqnarray}
V&=&\frac{1}{2}B_0\left[(m_u+m_d)(\pi^0)^2+\frac{1}{3}(m_u+m_d+4m_s)\eta^2\right]\nonumber\\
&&+\frac{B_0}{\sqrt{3}}(m_u-m_d)\pi^0\eta-C_4B_0^2F_\pi^3\left(\pi^0+\frac{1}{\sqrt{3}}\eta\right)\
,
\end{eqnarray}
which can be minimized to yield a condensate $\langle \pi^0\rangle$ and $\langle \eta \rangle$.

The above discussion can be easily generalized to an arbitrary four-quark operator, for which Eq. (\ref{linear}) can be written as
\begin{equation}\label{gpi}
{\cal L} = g_\pi C_4B_0^2F_\pi^3\pi^0+g_\eta C_4B_0^2F_\pi^3\eta +
... \ ,
\end{equation}
where $g_\pi$ and $g_\eta$ are numerical factors generated through the vacuum saturation approximation.
Then, the vevs of meson fields can be written as
\begin{eqnarray}\label{condensate}
\langle\pi^0\rangle&=&\frac{B_0F_\pi^3 C_4\left[g_\pi(m_u+m_d+4m_s)-\sqrt{3}g_\eta(m_u-m_d)\right]}{4(m_um_d+m_dm_s+m_sm_u)} \ , \nonumber\\
\langle\eta\rangle&=&\frac{B_0F_\pi^3
C_4\left[-\sqrt{3}g_\pi(m_u-m_d)+3g_\eta(m_u+m_d)\right]}{4(m_um_d+m_dm_s+m_sm_u)}
\ ,
\end{eqnarray}
which is inversely proportional to quark masses! The vev of $U$ can be written as
\begin{equation}
\langle
U\rangle=\exp\left[i\left(\begin{array}{ccc}\langle\pi^0\rangle+\frac{1}{\sqrt{3}}\langle\eta\rangle&0&0\\
0&-\langle\pi^0\rangle+\frac{1}{\sqrt{3}}\langle\eta\rangle&0\\
0&0&-\frac{2}{\sqrt{3}}\langle\eta\rangle\end{array}\right)/F_\pi\right]\
.
\end{equation}
This defines the vacuum state of Goldstone-boson fields.

Therefore, we can redefine the meson fields in the following way:
\begin{equation}
U=\langle U\rangle U'\ ,
\end{equation}
where $U'$ collects the physical meson excitations. Through this redefinition,
the meson lagrangian no longer contains terms annihilating the physical Goldstone
bosons. Correspondingly, we redefine the baryon fields,
\begin{equation}\label{cotransformation}
\xi B\xi=\langle U\rangle\xi' B'\xi'\ ,
\end{equation}
through a chiral transformation with $L=\langle U\rangle$ and $R=1$.

The above redefinition can change P-even and CP-even terms with explicit
chiral symmetry breaking to P-odd and CP-odd terms. This is particularly true
for the CP-even baryon lagrangian with linear dependence on quark masses,
\begin{equation}\label{cterm}
  {\cal L}_c = c_1{\rm Tr}[\bar BB]{\rm Tr}[M U^\dagger]+c_2{\rm
Tr}[M\xi^\dagger\bar BB\xi^\dagger]+c_3{\rm Tr}[\bar B\xi^\dagger
M\xi^\dagger B]\;+\;{\rm h.c.}
\end{equation}
and
\begin{eqnarray}\label{dterm}
{\cal L}_d = d_1{\rm Tr}[\bar B\gamma_5B]{\rm Tr}[M U^\dagger]+d_2{\rm
Tr}[M\xi^\dagger\bar B\gamma_5B\xi^\dagger]+d_3{\rm Tr}[\bar
B\gamma_5\xi^\dagger M\xi^\dagger B]\;+\;{\rm h.c.}
\end{eqnarray}
Substituting $\langle U\rangle$ to the above equation, we get CP-odd
pion-nucleon couplings through
\begin{equation}\label{76}
c_1{\rm Tr}[\bar B'B']{\rm Tr}[\langle U\rangle^\dagger M
U'^\dagger]+c_2{\rm Tr}[\langle U\rangle^\dagger M\xi'^\dagger\bar
B'B'\xi'^\dagger]+c_3{\rm Tr}[\bar B'\xi'^\dagger\langle
U\rangle^\dagger M\xi'^\dagger B']+{\rm h.c.}
\end{equation}
and the CP-odd masses of baryons
\begin{equation}\label{77}
d_1{\rm Tr}[\bar B'\gamma_5B']{\rm Tr}[\langle U\rangle^\dagger M
U'^\dagger]+d_2{\rm Tr}[\langle U\rangle^\dagger M\xi'^\dagger\bar
B'\gamma_5B'\xi'^\dagger]+d_3{\rm Tr}[\bar
B'\gamma_5\xi'^\dagger\langle U\rangle^\dagger M\xi'^\dagger
B']+{\rm h.c.}
\end{equation}
which is part of the CP-odd mass generated by the four-quark operator.
Note that since $\langle U\rangle$ is inversely proportional to
the quark mass, the above contribution is not suppressed in the
chiral limit.

One can also get a CP-odd dipole moment by considering a photo-pion
production term off the nucleon. When the pion is condensed through
CP-odd effects, one generates a new contribution to the CP-odd moment,
which is beyond the scope of this paper.

\subsection{Matching to CP-Odd Baryon Operators}

In this subsection, we construct the leading P-odd and CP-odd baryon
operators induced by the CP-odd four-quark operators. These include
all the operators with one baryon and one conjugate baryon fields,
and without any quark masses or derivatives. All the independent
operators are listed in Table \ref{Hadronic}.

\begin{table}[h]
\begin{tabular}{|c|c|}
\hline Rept.~&~Operators\\
\hline $(\bar 3,3)$~&~$\begin{array}{c}O_3^{(1)}=i{\rm Tr}[\bar
BB]{\rm Tr}[AU^\dagger], O_3^{(2)}=i{\rm Tr}[A\xi^\dagger\bar
BB\xi^\dagger], O_3^{(3)}=i{\rm Tr}[\bar B\xi^\dagger A\xi^\dagger
B],\\ \tilde O_3^{(1)}=i{\rm Tr}[\bar B\gamma_5B]{\rm
Tr}[AU^\dagger], \tilde O_3^{(2)}=i{\rm Tr}[A\xi^\dagger\bar
B\gamma_5B\xi^\dagger],
\tilde O_3^{(3)}=i{\rm Tr}[\bar B\gamma_5\xi^\dagger A\xi^\dagger B],\end{array}$\\
\hline $(6,\bar
6)$~&~$\begin{array}{c}O_6^{(1)}=iS^{ij}_{kl}(\xi\bar B\xi)^k_i(\xi
B\xi)^l_j,\; O_6^{(2)}=iS^{ij}_{kl}(\xi\bar BB\xi)^k_iU^l_j,\\
O_6^{(3)}=iS^{ij}_{kl}(\bar B\xi)^m_i(\xi B)^k_m U^l_j,\;
O_6^{(4)}=i{\rm Tr}[\bar BB]S^{ij}_{kl}U^k_iU^l_j.\\
\tilde O_6^{(1)}=iS^{ij}_{kl}(\xi\bar B\gamma_5\xi)^k_i(\xi
B\xi)^l_j,\; \tilde O_6^{(2)}=iS^{ij}_{kl}(\xi\bar B\gamma_5B\xi)^k_iU^l_j,\\
\tilde O_6^{(3)}=iS^{ij}_{kl}(\bar B\gamma_5\xi)^m_i(\xi B)^k_m
U^l_j,\;
\tilde O_6^{(4)}=i{\rm Tr}[\bar B\gamma_5B]S^{ij}_{kl}U^k_iU^l_j\\ \end{array}$\\
\hline $(8,8)$~&~$\begin{array}{c}O_8^{(1)}=i{\rm Tr}[\bar
B\xi^\dagger H_1 U H_2\xi^\dagger B],\;O_8^{(2)}=i{\rm
Tr}[\xi^\dagger\bar BB\xi^\dagger H_1 U H_2],\;\\O_8^{(3)}=i{\rm
Tr}[\xi\bar BB\xi H_2 U^\dagger H_1],\;O_8^{(4)}=i{\rm Tr}[\bar B\xi
H_2 U^\dagger H_1\xi B],\;\\O_8^{(5)}=i{\rm Tr}[\xi\bar B\xi^\dagger
H_1]{\rm Tr}[\xi^\dagger B\xi H_2],\;O_8^{(6)}=i{\rm
Tr}[\xi^\dagger\bar B\xi H_2]{\rm Tr}[\xi B\xi^\dagger
H_1],\;\\O_8^{(7)}=i{\rm Tr}[\xi \bar B\xi H_2\xi^\dagger
B\xi^\dagger  H_1],\;O_8^{(8)}=i{\rm Tr}[\xi^\dagger\bar
B\xi^\dagger H_1\xi B\xi H_2].\\ \tilde O_8^{(1)}=i{\rm Tr}[\bar
B\gamma_5\xi^\dagger H_1 U H_2\xi^\dagger B],\;\tilde
O_8^{(2)}=i{\rm
Tr}[\xi^\dagger\bar B\gamma_5B\xi^\dagger H_1 U H_2],\;\\
\tilde O_8^{(3)}=i{\rm Tr}[\xi\bar B\gamma_5B\xi H_2 U^\dagger
H_1],\;\tilde
O_8^{(4)}=i{\rm Tr}[\bar B\gamma_5\xi H_2 U^\dagger H_1\xi B],\;\\
\tilde O_8^{(5)}=i{\rm Tr}[\xi\bar B\gamma_5\xi^\dagger H_1]{\rm
Tr}[\xi^\dagger B\xi H_2],\;\tilde O_8^{(6)}=i{\rm
Tr}[\xi^\dagger\bar B\gamma_5\xi H_2]{\rm Tr}[\xi B\xi^\dagger H_1],\;\\
\tilde O_8^{(7)}=i{\rm Tr}[\xi \bar B\gamma_5\xi H_2\xi^\dagger
B\xi^\dagger H_1],\;\tilde O_8^{(8)}=i{\rm
Tr}[\xi^\dagger\bar
B\gamma_5\xi^\dagger H_1\xi B\xi H_2]\\ \end{array}$\\
\hline
\end{tabular}
\caption{Hadronic operators that have the same quantum numbers as
four-quark operators in different irreducible
representations.}\label{Hadronic}
\end{table}

There are two types of operators in Table \ref{Hadronic}, those
with and without tilde.
For the first group without tilde,
the expansion of the pion field generates the P-odd and
CP-odd nucleon-pion vertices
\begin{equation}
{\cal L}^{\rm CP-odd}_{NN\pi} = (h_c\bar p n\pi^+ +{\rm
h.c.})+h_n\bar n n\pi^0 + h_p \bar p p\pi^0\ .
\end{equation}
For the second group, the leading order expansion is a bilinear-baryon term
with a CP-odd mass structure,
\begin{equation}
{\cal L}_{\rm mass}^{\rm CP-odd}\sim-m_\star\bar ni\gamma_5 n.
\end{equation}
This term contributes to the CP-odd baryon wave function.

Traditionally, P-odd and CP-odd pion-nucleon couplings are defined
in terms of isospin 0, 1, and 2 of the operators, which can be
written as~\cite{Pospelov:2005pr}
\begin{equation}
{\cal L}_{\pi NN} = \bar g^{(0)}_{\pi NN}\bar N\tau^a N\pi^a + \bar
g^{(1)}_{\pi NN}\bar NN\phi^0 + \bar g^{(2)}_{\pi NN}(\bar
N\tau^aN\pi^a-3\bar N\tau^3N\pi^0)\ ,
\end{equation}
where $\bar g^{(i)}_{\pi NN}$ is the coupling of the isospin-$i$
term and $\tau^i$ are the Pauli matrices. Then, in terms of $\bar
g^{(i)}_{\pi NN}$, $h_c$, $h_n$, and $h_p$ can be written as
\begin{equation}
h_c = \sqrt{2}(\bar g^{(0)}_{\pi NN}+\bar g^{(2)}_{\pi NN}),\;\;h_n
= (-\bar g^{(0)}_{\pi NN}+\bar g^{(1)}_{\pi NN}+2\bar g^{(2)}_{\pi
NN}),\;\;h_p = (\bar g^{(0)}_{\pi NN}+\bar g^{(1)}_{\pi NN}-2\bar
g^{(2)}_{\pi NN})\ ,
\end{equation}
where $h_p$ does not contribute to nEDM.

To match the P-odd and CP-odd four-quark operators to the above
baryon operators, one must find ways to calculate the corresponding
non-perturbative Wilson coefficients. This can be done by
considering the matrix elements of the quark operators in simple
states. Take $O_{11}^{ud}=\bar ui\gamma_5u\bar dd$ as an example. As
shown in the last section,  it can be decomposed into irreducible
representations of the chiral group,
\begin{equation}\label{18}
O_{11}^{ud}\;=\;O_{11}^{ud,(\bar 3,3)}+O_{11}^{ud,(6,\bar
6)}+O_{11}^{ud,(8,8)} \;+\;{\rm h.c.}
\end{equation}
The spurions related to this operator are given in Eqs. (\ref{5}),
(\ref{6}) and (\ref{8}). $O_{11}^{ud,(\bar 3,3)}$,
$O_{11}^{ud,(6,\bar 6)}$ and $O_{11}^{ud,(8,8)}$ must be matched to
the hadronic operators in the same irreducible representations and
with the same spurions. Take the un-tilded hadronic operators as an
example:
\begin{eqnarray}\label{19'}
O^{ud,(\bar 3,3)}_{11}&=&\sum_{i=1}^3C_3^{(i)} O_3^{(i)} + ... \ , \nonumber\\
O^{ud,(6,\bar 6)}_{11}&=&\sum_{i=1}^4C_6^{(i)} O_6^{(i)} + ...\ , \nonumber\\
O^{ud,(8,8)}_{11}&=&\sum_{i=1}^8C_8^{(i)} O_8^{(i)} + ... \ ,
\end{eqnarray}
where ``...'' represents higher order operators.

Note that, an operator can be separated into hermitian part and
anti-hermitian part. Since the QCD Lagrangian is hermitian, the
hermitian part and the anti-hermitian part must have the same Wilson
coefficient in the effective theory. Take the operator $\bar q_L
q_R$ as an example, it is a $(\bar3,3)$ operator, so it can be
matched to $CU^\dagger$ in the chiral perturbation theory, while its
hermitian conjugation $\bar q_R q_L$ is matched to $CU$ with exactly
the same Wilson coefficient since the QCD Lagrangian is invariant
under the hermitian conjugate transformation. Therefore, the
hermitian part of $\bar q_L q_R$ can be matched to
$C(U^\dagger+U)/2$ whereas the anti-hermitian part can be matched to
$C(U^\dagger-U)/2$. As a result, one can use either the hermitian
part or the anti-hermitian part of the operators to get their Wilson
coefficients depending on which way is easier. For the operators
without tilde listed in Table \ref{Hadronic}, the anti-hermitian
parts contain terms having only one baryon field and one anti-baryon
field which is easy to do the matching, while for the operators with
tilde, the hermitian part is easier. Therefore, we choose to match
the anti-hermitian part of the operators without a tilde whereas
match the hermitian part of the operators with a tilde to get the
Wilson coefficients of them. One can show that this matching
procedure works when current algebra is valid such as in
non-relativistic quark model.

The leading-order expansion of the hadronic operators are, for
$(\bar 3 ,3)$ operators,
\begin{eqnarray}\label{19}
O_3^{(1)}&\simeq&i\bar p p+i\bar n
n+i\bar\Lambda\Lambda+i\bar\Sigma^0\Sigma^0+i\bar\Sigma^+\Sigma^+ +
i\bar\Sigma^-\Sigma^- + i\bar\Xi^0\Xi^0+ i\bar\Xi^-\Xi^-\ , \nonumber\\
O_3^{(2)}&\simeq&i\bar p p+i\bar n
n+\frac{2i}{3}\bar\Lambda\Lambda\ , \nonumber\\
O_3^{(3)}&\simeq&\frac{2i}{3}\bar\Lambda\Lambda+i\bar\Xi^0\Xi^0+i\bar\Xi^-\Xi^- \ .
\end{eqnarray}
Therefore, we can determine the Wilson coefficients with four physical
matrix elements,
\begin{eqnarray}\label{20}
C_3^{(1)}+C_3^{(2)}&=&(-i)\langle
p|O^{ud,(\bar3,3)}_{11}|p\rangle\ , \nonumber\\
C_3^{(1)}&=&(-i)\langle\Sigma^0|O^{ud,(\bar3,3)}_{11}|\Sigma^0\rangle\ , \nonumber\\
C_3^{(1)}+\frac{2}{3}C_3^{(2)}&=&(-i)\langle\Lambda|O^{ud,(\bar3,3)}_{11}|\Lambda\rangle\ , \nonumber\\
C_3^{(1)}+C_3^{(3)}&=&(-i)\langle\Xi^0|O^{ud,(\bar3,3)}_{11}|\Xi^0\rangle \ ,
\end{eqnarray}
where we have chosen the normalization condition
\begin{equation}
\langle \vec P|\vec P'\rangle = (2\pi)^3\delta^3(\vec P-\vec P')\ ,
\end{equation}
where $\vec P$ and $\vec P'$ are the momenta of the states.

 Since the number of equations is larger than the number of
variables, to get a solution the following condition must be
satisfied,
\begin{equation}
\det\left(\begin{array}{cccc}1&1&0&\langle
p|O^{ud,(\bar3,3)}_{11}|p\rangle\\
1&0&0&\langle\Sigma^0|O^{ud,(\bar3,3)}_{11}|\Sigma^0\rangle\\
1&\frac{2}{3}&0&\langle\Lambda|O^{ud,(\bar3,3)}_{11}|\Lambda\rangle\\
1&0&1&\langle\Xi^0|O^{ud,(\bar3,3)}_{11}|\Xi^0\rangle\\
\end{array}\right)=0 \ ,
\end{equation}
which gives a nontrivial relation among these matrix elements;
\begin{equation}
-\frac{2}{3}\langle
p|O^{ud,(\bar3,3)}|p\rangle-\frac{1}{3}\langle\Sigma^0|O^{ud,(\bar3,3)}_{11}|\Sigma^0\rangle+\langle\Lambda|O^{ud,(\bar3,3)}_{11}|\Lambda\rangle=0 \ .
\end{equation}
This relation must be satisfied in the chiral limit, so it is a test for direct calculations of the matrix elements.
Similarly, a simple inspection of  Eq. (\ref{19}) can give us some more
relations among matrix elements
\begin{eqnarray}
\langle p|O^{ud,(\bar3,3)}_{11}|p\rangle&=&\langle
n|O^{ud,(\bar3,3)}_{11}|n\rangle\ , \nonumber\\
\langle\Sigma^0|O^{ud,(\bar3,3)}_{11}|\Sigma^0\rangle&=&\langle\Sigma^+|O^{ud,(\bar3,3)}_{11}|\Sigma^+
\rangle\;=\;\langle\Sigma^-|O^{ud,(\bar3,3)}_{11}|\Sigma^-\rangle \ , \nonumber\\
\langle\Xi^0|O^{ud,(\bar3,3)}_{11}|\Xi^0\rangle&=&\langle\Xi^-|O^{ud,(\bar3,3)}_{11}|\Xi^-\rangle \ .
\end{eqnarray}

Generalizing the above discussion to $(6,\bar6)$ and $(8,8)$ operators,  we write down
the leading expansion of the hadronic operators,
\begin{eqnarray}\label{25}
O_6^{(1)}&\simeq&\frac{i}{3}\bar\Lambda\Lambda-i\bar\Sigma^0\Sigma^0+i\bar\Sigma^+\Sigma^++i\bar\Sigma^-\Sigma^- \ , \nonumber\\
O_6^{(2)}&\simeq&i\bar p p+i\bar n n+\frac{i}{3}\bar \Lambda\Lambda+i\bar\Sigma^0\Sigma^0+i\bar\Sigma^+\Sigma^++i\bar\Sigma^-\Sigma^-\ , \nonumber\\
O_6^{(3)}&\simeq&\frac{i}{3}\bar\Lambda\Lambda+i\bar\Xi^0\Xi^0+i\bar\Xi^-\Xi^-+i\bar\Sigma^0\Sigma^0+i\bar\Sigma^+\Sigma^++i\bar\Sigma^-\Sigma^- \ , \nonumber\\
O_6^{(4)}&\simeq&2i\left(\bar p p+\bar n
n+\bar\Lambda\Lambda+i\bar\Xi^0\Xi^0+\bar\Xi^-\Xi^-+\bar\Sigma^0\Sigma^0+\bar\Sigma^+\Sigma^++\bar\Sigma^-\Sigma^-\right)\;;
\end{eqnarray}

\begin{eqnarray}
O_8^{(1)}&\simeq&i\left(\bar p
p+\frac{1}{6}\bar\Lambda\Lambda+\frac{1}{2\sqrt{3}}\bar\Lambda\Sigma^0+\frac{1}{2\sqrt{3}}\bar\Sigma^0\Lambda
+\frac{1}{2}\bar\Sigma^0\Sigma^0+\bar\Sigma^+\Sigma^+\right)\ , \nonumber\\
O_8^{(2)}&\simeq&i\left(\frac{1}{6}\bar\Lambda\Lambda+\bar\Xi^-\Xi^-+\frac{1}{2\sqrt{3}}\bar\Lambda\Sigma^0+\frac{1}{2\sqrt{3}}\bar\Sigma^0\lambda+
\frac{1}{2}\bar\Sigma^0\Sigma^0+\bar\Sigma^-\Sigma^-\right)\ , \nonumber\\
O_8^{(3)}&\simeq&i\left(\frac{1}{6}\bar\Lambda\Lambda+\bar\Xi^0\Xi^0-\frac{1}{2\sqrt{3}}\bar\Lambda\Sigma^0-\frac{1}{2\sqrt{3}}\bar\Sigma^0\lambda+
\frac{1}{2}\bar\Sigma^0\Sigma^0+\bar\Sigma^+\Sigma^+\right)\ , \nonumber\\
O_8^{(4)}&\simeq&i\left(\bar n
n+\frac{1}{6}\bar\Lambda\Lambda-\frac{1}{2\sqrt{3}}\bar\Lambda\Sigma^0-\frac{1}{2\sqrt{3}}\bar\Sigma^0\Lambda+\frac{1}{2}\bar\Sigma^0\Sigma^0+\bar\Sigma^-\Sigma^-\right)\nonumber\\
O_8^{(5)}&\simeq&i\bar\Sigma^+\Sigma^+\ , \nonumber\\
O_8^{(6)}&\simeq&i\bar\Sigma^-\Sigma^-\ , \nonumber\\
O_8^{(7)}&\simeq&i\left(\frac{1}{6}\bar\Lambda\Lambda+\frac{1}{2\sqrt{3}}\bar
\Lambda\Sigma^0-\bar\Sigma^0\Lambda-\frac{1}{2}\bar\Sigma^0\Sigma^0\right)\ , \nonumber\\
O_8^{(8)}&\simeq&i\left(\frac{1}{6}\bar\Lambda\Lambda-\frac{1}{2\sqrt{3}}\bar
\Lambda\Sigma^0+\bar\Sigma^0\Lambda-\frac{1}{2}\bar\Sigma^0\Sigma^0\right)\;,
\end{eqnarray}
from which we can get similar relations among matrix elements just
like in the $(\bar3,3)$ case shown in Table \ref{relations}. The
other four-quark operators with the same flavor structures have the
same relations among hadronic matrix elements as in this case.

\begin{table}[h]
\begin{tabular}{|c|c|}
\hline Rep.~&~Relations\\
\hline $(\bar 3,3)$~&~$\begin{array}{c}\langle
p|O^{ud,(\bar3,3)}_{11}|p\rangle=\langle
n|O^{ud,(\bar3,3)}_{11}|n\rangle,\langle\Xi^0|O^{ud,(\bar3,3)}_{11}|\Xi^0\rangle=\langle\Xi^-|O^{ud,(\bar3,3)}_{11}|\Xi^-\rangle,\\
\langle\Sigma^0|O^{ud,(\bar3,3)}_{11}|\Sigma^0\rangle=\langle\Sigma^+|O^{ud,(\bar3,3)}_{11}|\Sigma^+\rangle=\langle\Sigma^-|O^{ud,(\bar3,3)}_{11}|\Sigma^-\rangle,\\
-\frac{2}{3}\langle
p|O^{ud,(\bar3,3)}_{11}|p\rangle-\frac{1}{3}\langle\Sigma^0|O^{ud,(\bar3,3)}_{11}|\Sigma^0\rangle+\langle\Lambda|O^{ud,(\bar3,3)}_{11}|\Lambda\rangle=0\\
\end{array}$\\
\hline $(6,\bar 6)$~&~$\begin{array}{c}\langle
p|O^{ud,(6,\bar6)}_{11}|p\rangle=\langle
n|O^{ud,(6,\bar6)}_{11}|n\rangle,\langle\Sigma^+|O^{ud,(6,\bar6)}_{11}|\Sigma^+\rangle=\langle\Sigma^-|O^{ud,(6,\bar6)}_{11}|\Sigma^-\rangle,
\\
\langle\Xi^+|O^{ud,(6,\bar6)}_{11}|\Xi^+\rangle = \langle\Xi^-|O^{ud,(6,\bar6)}_{11}|\Xi^-\rangle\\
2\langle p|O^{ud,6,\bar6}_{11}|p\rangle-\langle\Sigma^0|O^{ud,(6,\bar6))}_{11}|\Sigma^0\rangle-3\langle\Lambda|O^{ud,(6,\bar6)}_{11}|\Lambda\rangle+2\langle\Xi^0|O^{ud,(6,\bar6)}_{11}|\Xi^0\rangle=0\\ \end{array}$\\
\hline $(8,8)$~&~$\begin{array}{c}\langle p|O^{ud,(8,8)}_{11}|p\rangle+\langle n|O^{ud,(8,8)}_{11}|n\rangle+\langle \Sigma^0|O^{ud,(8,8)}_{11}|\Sigma^0\rangle-3\langle \Lambda|O^{ud,(8,8)}_{11}|\Lambda\rangle\\+\langle \Xi^0|O^{ud,(8,8)}_{11}|\Xi^0\rangle+\langle \Xi^-|O^{ud,(8,8)}_{11}|\Xi^-\rangle=0\\
\langle p|O^{ud,(8,8)}_{11}|p\rangle-\langle n|O^{ud,(8,8)}_{11}|n\rangle-\langle \Xi^0|O^{ud,(8,8)}_{11}|\Xi^0\rangle+\langle \Xi^-|O^{ud,(8,8)}_{11}|\Xi^-\rangle\\
-\sqrt{3}\langle\Lambda|O^{ud,(8,8)}_{11}|\Sigma^0\rangle-\sqrt{3}\langle\Sigma^0|O^{ud,(8,8)}_{11}|\Lambda\rangle=0\\
\end{array}$\\
\hline
\end{tabular}
\caption{Relations among hadronic matrix elements of the four-quark operators in different
chiral representations.}\label{relations}
\end{table}

One can either build models or do lattice QCD calculations to get
these simplest four-quark matrix elements. Once known, one can get
the Wilson coefficients by solving Eq. (\ref{20}) and similar
equations for $(6,\bar6)$ and $(8,8)$ operators. Then one can expand
these hadronic operators to the first order with one meson field in
each term to get the P-odd and CP-odd pion-nucleon vertices. A
similar method works for baryon operators with tilde. We will
consider these matrix elements in the next section.

\subsection{Matching to EDM-Type Operators}

In $\chi$PT, any CP-odd quark-gluon operator will generate directly
an EDM contribution to the neutron, analytical in the chiral limit.
To write down such a contribution, introduce vector and axial vector
octet potential $v_\mu$ and $a_\mu$, which transform under local
chiral rotations (with space-time dependent chiral transformation)
as
\begin{eqnarray}
   r_\mu \equiv v_\mu+a_\mu \longrightarrow v_\mu' + a_\mu' &=& R(v_\mu+a_\mu)R^\dagger + iR\partial_\mu R^\dagger \ , \nonumber \\
   l_\mu \equiv v_\mu-a_\mu \longrightarrow v_\mu' - a_\mu' &=& L(v_\mu-a_\mu)L^\dagger + iL\partial_\mu L^\dagger \ .
\end{eqnarray}
The corresponding gauge fields are defined as
\begin{eqnarray}
    f^R_{\mu\nu} &=& \partial_\mu r_\nu - \partial_\nu r_\mu - i[r_\mu,r_\nu] \ , \nonumber \\
     f^L_{\mu\nu} &=& \partial_\mu l_\nu - \partial_\nu l_\mu - i[l_\mu,l_\nu] \ .
\end{eqnarray}
The gauge fields with definite parity are defined as
\begin{eqnarray}
      f^{\pm}_{\mu\nu} = \xi^\dagger f^R_{\mu\nu} \xi \pm  \xi f^L_{\mu\nu} \xi^\dagger ,
\end{eqnarray}
which transform under chiral transformation as
\begin{equation}
     f^{\pm}_{\mu\nu}   \rightarrow K f^{\pm}_{\mu\nu} K^\dagger \ .
\end{equation}
When reducing to the electromagnetic field, $a_\mu=0$,
$f^{\pm}_{\mu\nu} = (\xi^\dagger Q \xi \pm \xi Q\xi^\dagger)
F_{\mu\nu}$, where $Q=diag(2/3,-1/3,-1/3)$ and $F_{\mu\nu}$ is the
electromagnetic field~\cite{Bernard:1995dp}.

One can write down a number of EDM type of operators which contain
$\bar B$ and $B$, $f^{\pm}_{\mu\nu}$, and the spurion fields $A$,
$H$, and $S$. These contributions are direct matching contributions
to the neutron EDM, and cannot be calculated in $\chi$PT. These
chiral constants can in principle be calculated in lattice QCD.
However, we will present quark-model estimates in Sec. V.



\subsection{Peccei-Quinn Symmetry and Induced $\theta$-Term}

The experimental upper bound on the neutron EDM gives a strong
constraint on the P-odd and CP-odd $\theta$-term, $\theta G\tilde G$,
in the QCD lagrangian \cite{Crewther:1979pi,Pospelov:2000bw,Baluni:1978rf}.
Using the current experimental limit \cite{Baker:2006ts},
\begin{equation}
d_n<2.9\times10^{-26} e\;{\rm cm} \ ,
\end{equation}
one can get the upper bound,
\begin{equation}
\theta<10^{-10} \ .
\end{equation}
On the other hand, it is unnatural for a parameter of the
fundamental theory to be so small without fine tuning. There are
generally two ways to solve this strong CP problem in the
literature. The first is by introducing the spontaneous breaking of
parity. Since the $\theta$-term also breaks parity, if at some high
energy scale parity is conserved, then the $\theta$-term at low
energy scale can only be generated by loop effects and will be
suppressed naturally \cite{Babu:2001se}.

The other way is to introduce the Peccei-Quinn
symmetry, $U(1)_A$ \cite{Peccei:1977hh}. After the spontaneous breaking of
the symmetry, there emerges a pseudo-goldstone boson, $a$,
which is called the axion~\cite{Weinberg:1975ui,Kim:1979if,Shifman:1979if}.
The effective Lagrangian for the axion field can be written as
\begin{equation}
{\cal L}_{a}=\frac{1}{2}\partial_\mu a\partial^\mu
a+\frac{a}{f_a}\frac{\alpha_s}{8\pi}G^a_{\mu\nu}\tilde G^{a\mu\nu},
\end{equation}
which includes an effective interaction with $G\tilde G$.
The axion field gets a small mass through the non-vanishing correlation
function
\begin{equation}
K=i\left\{\int d^4x e^{ik\cdot
x}\left\langle0\left|T\left(\frac{\alpha_s}{8\pi}G\tilde
G(x),\frac{\alpha_s}{8\pi}G\tilde
G(0)\right)\right|0\right\rangle\right\}_{k=0} \ ,
\end{equation}
after taking into account the non-perturbative QCD
effect~\cite{Kim:1979if,Shifman:1979if}.

When there is an additional neutral P-odd, CP-odd quark operator,
$O_{\rm CP-odd}$, in the lagrangian, the correlation function
\begin{equation}
K_1=i\left\{\int d^4x e^{ik\cdot
x}\left\langle0\left|T\left(\frac{\alpha_s}{8\pi}G\tilde G(x),O_{\rm
CP-odd}(0)\right)\right|0\right\rangle\right\}_{k=0}
\end{equation}
will be generally nonzero. Therefore, the vev of $a$, which cancels
precisely the $\theta$-term in the original lagrangian, will now be
shifted by a small amount proportional to $K_1$. A non-vanishing
effective $\theta$-term is induced as \cite{Falk:1999tm}
\begin{equation}
\theta_{\rm ind}=-\frac{K_1}{K} \ ,
\end{equation}
which can contribute to the neutron EDM.

Following Ref. \cite{Falk:1999tm}, we take the operator $\bar
ui\gamma_5u\bar d d$ as an example to calculate the contribution to
neutron EDM through the induced $\theta$-term. Then, $K_1$ can then
be written as
\begin{equation}
K_1=i\left\{\int d^4x e^{ik\cdot
x}\left\langle0\left|T\left(\frac{\alpha_s}{8\pi}G\tilde
G(x),C_4\bar ui\gamma_5 u\bar d
d(0)\right)\right|0\right\rangle\right\}_{k=0}.
\end{equation}
Using the chiral anomaly \cite{Adler:1969gk}, one can get
\begin{eqnarray}
\frac{\alpha_s}{4\pi}G\tilde G&=&\partial_\mu J_5^\mu-2m_*(\bar u
i\gamma_5 u+\bar d i\gamma_5 d + \bar si\gamma_5 s),
\end{eqnarray}
where
\begin{equation}\label{J-5}
J_5^\mu\equiv\left(\frac{m_*}{m_u}\bar
u\gamma^\mu\gamma_5u+\frac{m_*}{m_d}\bar d \gamma^\mu\gamma_5
d+\frac{m_*}{m_s}\bar s\gamma^\mu\gamma_5s\right).
\end{equation}
Then one can get
\begin{eqnarray}
K_1&=&\frac{i}{2}\int d^4x e^{ik\cdot
x}\left\langle0\left|T\left(\partial_\mu J_5^\mu(x),C_4\bar
ui\gamma_5u\bar d d(0)\right)\right|0\right\rangle_{k=0}\nonumber\\
&&-\frac{i}{2}\int d^4x e^{ik\cdot
x}\left\langle0\left|T\left(2m_*(\bar u i\gamma_5 u+\bar d i\gamma_5
d + \bar si\gamma_5 s)(x),C_4\bar ui\gamma_5u\bar d
d(0)\right)\right|0\right\rangle_{k=0}.
\end{eqnarray}
The second term on the right-hand side of the above equation is
negligible compared to the first term because it is explicitly
proportional to the reduced quark mass $m_*$ and the operator $\bar
ui\gamma_5u+\bar di\gamma_5d+\bar si\gamma_5s$ cannot annihilate
light mesons. Therefore $K_1$ can be calculated as
\begin{eqnarray}
K_1&\approx&\frac{i}{2}\int d^4x e^{ik\cdot
x}\langle0|T(\partial_\mu
J^\mu_5(x),C_4\bar ui\gamma_5u\bar d d(0))|0\rangle_{k=0}\nonumber\\
&=&-\frac{i}{2}C_4\langle0|[Q_5(0),\bar ui\gamma_5u\bar d
d(0)]|0\rangle,
\end{eqnarray}
where $Q_5$ is the charge related to the current $J_5^\mu$ defined
in Eq. (\ref{J-5}).  In the spirit of large $N_C$
\cite{'tHooft:1973jz,Manohar:1998xv} expansion one can assume that
\begin{equation}
\langle0|\bar u i\gamma_5 u\bar di\gamma_5 d|0\rangle\ll\langle
0|\bar u u \bar d d|0\rangle\approx\langle0|\bar u
u|0\rangle\langle0|\bar d d|0\rangle.
\end{equation}
Therefore, we can get
\begin{eqnarray}
K_1&\simeq&-\frac{i}{2}C_4\langle0|[Q_5(0),\bar
ui\gamma_5u(0)]|0\rangle\langle0|\bar d
d|0\rangle=-C_4\frac{m_*}{m_u}\langle0|\bar u
u|0\rangle\langle0|\bar d
d|0\rangle\nonumber\\
&=&-\frac{m_*}{m_u}C_4B_0^2F_\pi^4 \ .
\end{eqnarray}
Using the previously known result \cite{Shifman:1979if}
\begin{equation}
K=-m_*F_\pi^2 B_0 \ ,
\end{equation}
one can get the $\theta$ angle induced by the operator
$\bar u i\gamma_5 u\bar d d $,
\begin{equation}
\theta_{\rm ind}=-\frac{K_1}{K}=-\frac{C_4B_0F_\pi^2}{m_u}\;.
\end{equation}
A similar result can be obtained for any other CP-odd
four-quark operator.

Using the standard chiral result in the
literature~\cite{Pich:1991fq}, we write down the effective chiral
lagrangian corresponding to this induced $\theta$ term;
\begin{equation}\label{51}
{\cal L}_{\theta}=\frac{4\theta m_*}{F_\pi}\left(c_2{\rm
Tr}[\Sigma\bar BB]+c_3{\rm Tr}[\bar B\Sigma
B]\right)+2m_*\theta(3d_1+d_2+d_3){\rm Tr}[\bar Bi\gamma_5B]\ .
\end{equation}
From the above, we read off the CP-odd pion-nucleon coupling and the
CP-odd mass of the neutron;
\begin{eqnarray}
&&h_c=-\frac{2\sqrt{2}C_4B_0F_\pi
m_*}{m_u}\;\;\;h_n=\frac{2C_4B_0F_\pi m_*}{m_u}\;,\nonumber\\
&& M_\star = \frac{2C_4B_0F_\pi^2 m_*}{m_u}(3d_1+d_2+d_3)\ .
\end{eqnarray}
Comparing this with the meson condensates contribution in Eq.
(\ref{90}), one finds that they are in the same order. If the
Peccei-Quinn symmetry exists, one should add this contribution to
the neutron EDM. However, since it is not known if the axion
mechanism is in operation, we will not include this contribution to
the nEDM in the remainder of the paper.

\section{P-odd and CP-odd nucleon-pion vertices and CP-odd Nucleon Mass}

In this section, we study the induced physical P-odd and CP-odd
nucleon-pion vertices as well as the CP-odd nucleon mass from
four-quark operators.  There are a number of contributions to
consider: First, the CP-odd meson lagrangian will generate meson
condensates which can convert a CP-even vertex into a CP-odd one.
Second, the baryon wave function contains the CP-odd part due to the
CP-odd nucleon mass, which can also rotate a CP-even coupling into a
CP-odd one. Finally, there is the contribution from the direct
matching operators (without a tilde) in TABLE II. We will consider
all of these in this section.

\subsection{Meson Condensates Contribution}

We use the vacuum saturation approximation to calculate the meson
effective lagrangian; the vevs of $\pi^0$ and $\eta$ can be obtained
from Eq. (\ref{condensate}), where $g_\pi$ and $g_\eta$ for all the
four-quark operators built with color-singlet and octet scalar
currents are listed in Table \ref{gpigeta}. Those induced by tensor
operators vanish in this approximation.

\begin{table}[h]
\begin{tabular}{|c|c|c|c|c|c|}
\hline Operator~&~$g_\pi$~&~$g_\eta$&Operator~&~$g_\pi$~&~$g_\eta$\\
\hline $\bar ui\gamma_5u\bar d d$~&~$1$~&~$1/\sqrt{3}$&$\bar
ui\gamma_5 t^a u\bar d t^a d$~&~$0$~&~$0$\\
\hline $\bar di\gamma_5d\bar u u$~&~$-1$~&~$1/\sqrt{3}$&$\bar
di\gamma_5 t^a d\bar u  t^a u$~&~$0$~&~$0$\\
\hline $\bar ui\gamma_5u\bar s s$~&~$1$~&~$1/\sqrt{3}$&$\bar ui\gamma_5 t^a u\bar s  t^a s$~&~$0$~&~$0$\\
\hline $\bar si\gamma_5s\bar u u$~&~$0$~&~$-2/\sqrt{3}$&$\bar si\gamma_5 t^a s\bar u  t^a u$~&~$0$~&~$0$\\
\hline $\bar di\gamma_5d\bar s s$~&~$-1$~&~$1/\sqrt{3}$&$\bar di\gamma_5 t^a d\bar s  t^a s$~&~$0$~&~$0$\\
\hline $\bar si\gamma_5s\bar d d$~&~$0$~&~$-2/\sqrt{3}$&$\bar si\gamma_5 t^a s\bar d  t^a d$~&~$0$~&~$0$\\
\hline $\bar ui\gamma_5u\bar u u$~&~$5/6$~&~$5/(6\sqrt{3})$&$\bar ui\gamma_5 t^a u\bar u  t^a u$~&~$-2/9$~&~$-2/(9\sqrt{3})$\\
\hline $\bar di\gamma_5d\bar d d$~&~$-5/6$~&~$5/(6\sqrt{3})$&$\bar di\gamma_5 t^a d\bar d  t^a d$~&~$2/9$~&~$-2/(9\sqrt{3})$\\
\hline $\bar si\gamma_5s\bar s s$~&~$0$~&~$-5/(3\sqrt{3})$&$\bar
si\gamma_5 t^a s\bar s  t^a s$~&~$0$~&~$4/(9\sqrt{3})$\\ \hline
\end{tabular}
\caption{$g_\pi$ and $g_\eta$ induced by four-quark operators
constructed by scalar currents. Those induced by products of tensor currents are zero.}\label{gpigeta}
\end{table}

\begin{figure}[hbt]
\begin{center}
\includegraphics[width=10cm]{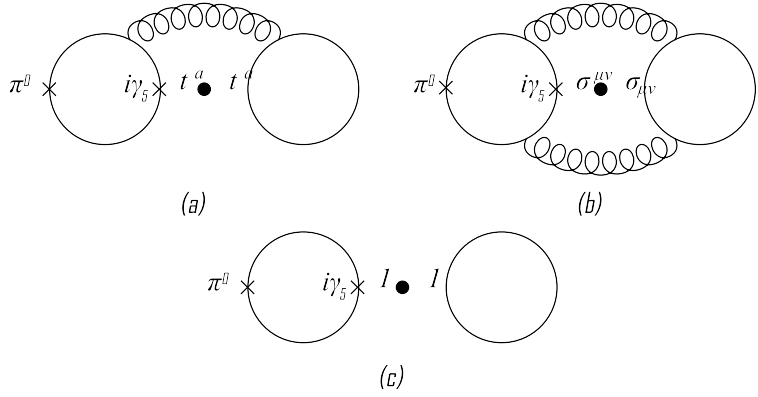}
\caption{Annihilation of pion by four-quark operators: (a)
operator constructed from two color-octet current, like
$\bar u i\gamma_5 t^a u \bar d t^a d$; (b) operator from two tensor currents, like $\bar
u\sigma^{\mu\nu}i\gamma_5 u\bar d\sigma_{\mu\nu}d$; (c)
operator from two scalar currents, like $\bar u i\gamma_5
u \bar d d$.}\label{largen}
\end{center}
\end{figure}

In the large $N_c$ QCD \cite{'tHooft:1973jz} (also see Ref.
\cite{Manohar:1998xv} for a good review), the leading contributions for operators
constructed from two color-octet currents
and two tensor currents are shown as diagrams (a) and (b) in Fig.
\ref{largen}, respectively. Detailed analysis shows that the diagrams
(a) and (b) suffer from $1/N_c^2$ suppressions compared with
(c), which stands for the operator constructed from two scalar
color-singlet currents.

\begin{table}[h]
\begin{tabular}{|c|c|c|c|c|}
\hline  & \multicolumn{2}{c|}{ Meson condensates contribution} & \multicolumn{2}{c|}{ Factorization}\\
\cline{2-5} Operator~&~$h_c$ ($C_4 B_0^2$)~&~$h_n$ ($C_4 B_0^2$)~&~$h_c$ ($C_4 B_0^2$)~&~$h_n$ ($C_4 B_0^2$)\\
\hline $\bar ui\gamma_5u\bar d d$~&~$-0.0117$~&~$0.225$~&~$0.0063 $~&~$-0.24 $ \\
\hline $\bar di\gamma_5d\bar u u$~&~$0.0130$~&~$-0.227$~&~$0.0063 $~&~$0.19 $\\
\hline $\bar ui\gamma_5u\bar s s$~&~$-0.0117$~&~$0.225$~&~$ 0$~&~$-0.088 $\\
\hline $\bar si\gamma_5s\bar u u$~&~$-0.00122$~&~$0.000864$~&~$0 $~&~$0 $\\
\hline $\bar di\gamma_5d\bar s s$~&~$0.0130$~&~$-0.227$~&~$0 $~&~$0.087 $\\
\hline $\bar si\gamma_5s\bar d d$~&~$-0.00122$~&~$0.000864$~&~$0 $~&~$0 $\\
\hline $\bar ui\gamma_5u\bar u u$~&~$-0.00976$~&~$0.188$~&~$0 $~&~$-0.16 $\\
\hline $\bar di\gamma_5d\bar d d$~&~$0.0108$~&~$-0.189$~&~$0 $~&~$0.20 $\\
\hline $\bar si\gamma_5s\bar s s$~&~$-0.00102$~&~$0.000722$~&~$0 $~&~$0 $\\
\hline $\bar ui\gamma_5\sigma^{\mu\nu}u\bar
d\sigma_{\mu\nu}d$~&~$0$~&~$0$~&~$0.076$~&~$0$\\
\hline $\bar ui\gamma_5\sigma^{\mu\nu}u\bar
s\sigma_{\mu\nu}s$~&~$0$~&~$0$~&~$0$~&~$0$\\
\hline $\bar di\gamma_5\sigma^{\mu\nu}d\bar
s\sigma_{\mu\nu}s$~&~$0$~&~$0$~&~$0$~&~$0$\\
\hline
\hline $\bar ui\gamma_5 t^a u\bar d t^a d$~&~$0$~&~$0$~&~$0.0085 $~&~$0 $\\
\hline $\bar di\gamma_5 t^a d\bar u  t^a u$~&~$0$~&~$0$~&~$0.0085 $~&~$0 $\\
\hline $\bar ui\gamma_5 t^a u\bar s  t^a s$~&~$0$~&~$0$~&~$0 $~&~$0 $\\
\hline $\bar si\gamma_5 t^a s\bar u  t^a u$~&~$0$~&~$0$~&~$0 $~&~$0 $\\
\hline $\bar di\gamma_5 t^a d\bar s  t^a s$~&~$0$~&~$0$~&~$0 $~&~$0 $\\
\hline $\bar si\gamma_5 t^a s\bar d  t^a d$~&~$0$~&~$0$~&~$0 $~&~$0 $\\
\hline $\bar ui\gamma_5 t^a u\bar u  t^a u$~&~$0.00261$~&~$-0.0501$~&~$0 $~&~$0.042 $\\
\hline $\bar di\gamma_5 t^a d\bar d  t^a d$~&~$-0.00288$~&~$0.0503$~&~$0 $~&~$-0.054 $\\
\hline $\bar si\gamma_5 t^a s\bar s  t^a s$~&~\;\;\;\;\;$0.000272\;\;\;\;\;$~&~$-0.000192$~&~$0 $~&~$0 $\\
\hline $\bar ui\gamma_5\sigma^{\mu\nu}t^au\bar
d\sigma_{\mu\nu}t^ad$~&~$0$~&~$0$~&~$0.101$~&~$0$\\
\hline $\bar ui\gamma_5\sigma^{\mu\nu}t^au\bar
s\sigma_{\mu\nu}t^as$~&~$0$~&~$0$~&~$0$~&~$0$\\
\hline $\bar di\gamma_5\sigma^{\mu\nu}t^ad\bar
s\sigma_{\mu\nu}t^as$~&~$0$~&~$0$~&~$0$~&~$0$\\
\hline
\end{tabular}
\caption{CP-odd pion-nucleon couplings induced by meson condensates.
$C_4$ is the Wilson coefficient of the corresponding four-quark
operator. The two columns on the right side shows the P-odd and
CP-odd pion-nucleon vertices calculated using factorization which
will be discussed in Sec. V. }\label{vevinducedh}
\end{table}

Terms contributing to the P-odd, CP-odd nucleon-pion vertices
through the condensates of neutral mesons are shown in Eq.
(\ref{76}). At tree level, one can relate the coefficients $c_1$,
$c_2$, and $c_3$ to the mass differences of the baryons and the $\pi
N$ $\sigma$-term, and their values can be found in the literature
\cite{Bernard:1995dp};
\begin{equation}\label{cs}
c_1=2 B_0 b_0\;,\;\;c_2=2 B_0 (b_d-b_f)\;,\;\;c_3=2 B_0 (b_d +
b_f)\;,
\end{equation}
where $b_0 = -0.517$ GeV$^{-1}$, $b_d = 0.066$ GeV$^{-1}$ and $b_f =
-0.213$ GeV$^{-1}$.

The vertices we are interested in have two nucleons and one pion
because of the infrared enhancement in the pion loop
\cite{Crewther:1979pi}. From Eq. (\ref{76}) we can read off the
relevant terms,
\begin{eqnarray}\label{90}
&&-\frac{1}{3F_\pi^2}\left\{c_3[3\sqrt{2}\langle\pi^0\rangle(m_u-m_d)+\sqrt{6}\langle\eta\rangle(m_u+m_d)]\right\}(\bar
n p \pi^- + \bar p n \pi^+)\nonumber\\
&&-\frac{2}{3F_\pi^2}\left\{c_3m_d(3\langle\pi^0\rangle-\sqrt{3}\langle\eta\rangle)+c_1[3(m_u+m_d)\langle\pi^0\rangle+\sqrt{3}\langle\eta\rangle(m_u-m_d)]\right\}\bar
n n\pi^0 \ ,
\end{eqnarray}
in which $\langle\pi^0\rangle$ and $\langle\eta\rangle$ are given in
Eq. (\ref{condensate}). It is customary to
define the P-odd, CP-odd nucleon-pion couplings
\begin{equation}\label{91}
{\cal L}_{CPV}=h_c (\bar p n \pi^+ + \bar n p \pi^-) + h_n \bar n n
\pi^0,
\end{equation}
where $h_c$ and $h_n$ induced by meson condensates are listed in
Table \ref{vevinducedh}. Typical values of $h_c$ are one order of
magnitude smaller than the value of $h_n$ because
$\sqrt{2}c_3(m_d-m_u) \ll 4c_1(m_u+m_d) $. For $h_c$ or $h_n$
generated by a certain four-quark operator, if the contribution from
$\langle\pi^0\rangle$ is non-vanishing, the contribution from
$\langle\eta\rangle$ can be neglected since
$\langle\pi^0\rangle/\langle\eta\rangle\simeq m_s/\hat m\simeq30$.
This also explains the contributions from operators with the $\bar
si\gamma_5 s$ factor are much smaller than those without. Finally,
the contributions from operators made of color-octet currents are
smaller than those from operators made of color-singlet currents
because a Fierz transformation is needed for color-octet operators
to annihilate the mesons, introducing a suppressing factor of 1/4.

In Table \ref{vevinducedh} one can see that the P-odd and CP-odd
pion-nucleon couplings are proportional to $B_0^2$, which is related
to the quark condensates. The value of $B_0$ can be extracted from
the pion mass
\begin{equation}
m_\pi^2 = B_0 (m_u + m_d)\;.
\end{equation}
The natural scale for $\chi$PT is $4\pi
F_\pi$~\cite{Manohar:1983md}, and for simplicity we use the same scale to define
the quark masses to get $B_0$. The quark masses we use are
$m_u=2.4$ MeV and $m_d=4.75$ MeV in $\overline{\rm MS}$ at 2 GeV.
Using the one-loop renormalization group to run them down to $\mu=4\pi F_\pi$, we have
\begin{equation}
B_0 = 2.2~ {\rm GeV}\;.
\end{equation}
Here we have used one-loop $\Lambda_{\rm QCD}= 250$ MeV.

\subsection{Direct Contribution from Matching}

To get the P-odd and CP-odd meson-nucleon coupling through direct
matching, one needs to calculate the matrix elements listed in Table
\ref{relations}. Lattice QCD is perhaps the ultimate choice for
calculating hadronic matrix elements. However, it is still quite
difficult to directly calculate the matrix elements of four-quark
operators between baryons. Therefore, we resort to quark models to
get an estimate. In the remainder of this subsection we will use two
different quark models to calculate these hadronic matrix elements:
the simple non-relativistic quark
model~\cite{Greenberg:1964pe,Faiman:1968js,Donoghue:1981sh,Koniuk:1979vy,Donoghue:1992dd}
and the MIT bag
model~\cite{Chodos:1974je,Johnson:1978uy,Chodos:1974pn,DeGrand:1975cf,Donoghue:1992dd}.
We also discuss the significance of the model calculations from the
viewpoint of naive factorization.

\subsubsection{Non-relativistic Quark Model}

Here we consider the simplest version of the non-relativistic quark
model with harmonic oscillator interacting potentials,
\begin{equation}
H=-\sum_{i=1}^3\frac{1}{2m}\nabla_i^2+\frac{1}{2}\frac{m_c}{3}\omega^2\left[(\vec
r_1-\vec r_2)^2+(\vec r_2-\vec r_3)^2+(\vec r_3-\vec r_1)^2\right]\
,
\end{equation}
where $\vec r_1$, $\vec r_2$, and $\vec r_3$ are positions of the
three quarks inside the baryon, $m_c$ is the mass of the constituent
quarks and $\omega$ is the angular frequency. One can isolate the
center of mass by introducing the Jacobi coordinates,
\begin{eqnarray}
\vec R&=&\frac{1}{\sqrt{3}}(\vec r_1 + \vec r_2 + \vec r_3) \ ,
\nonumber\\
\vec \rho &=&\frac{1}{\sqrt{2}}(\vec r_1 - \vec r_2) \ , \nonumber\\
\vec \lambda &=& \frac{1}{\sqrt{6}}(\vec r_1 + \vec r_2 -2\vec r_3) \ .
\end{eqnarray}
Then the spatial wave function of the nucleon can be written as
\begin{equation}\label{55}
f(\vec r_1,\vec r_2,\vec r_3;\vec
k)=(3\sqrt{3})^{-1/2}\exp(i\vec P\cdot \vec
R/\sqrt{3})\psi(\vec\rho,\vec\lambda) \ ,
\end{equation}
where $ \psi(\vec\rho,
\vec\lambda)=({\alpha^3}/{\pi^{3/2}})\exp\left[-\alpha^2(\rho^2+\lambda^2)/2\right]$
in which $\alpha = (m\omega)^{1/2}\approx0.41$ GeV
\cite{Koniuk:1979vy} is the oscillator parameter, and $\vec P$ is
the nucleon momentum. It is easy to check that the wave function is
normalized to $(2\pi)^3\delta^3(\vec P-\vec P')$. The internal part
of the wave function is assumed to have $SU(6)$ spin-flavor
symmetry. For example, the spin-up proton state has the following
wave function,
\begin{eqnarray}\label{95}
|p_\uparrow\rangle&=&\frac{1}{\sqrt{18}}\int d^3r_1 d^3r_2 d^3r_3
f(\vec r_1,\vec r_2,\vec
r_3)\epsilon^{abc}\left[u^{a\dagger}_\downarrow(\vec
r_1)d_\uparrow^{b\dagger}(\vec r_2)-u^{a\dagger}_\uparrow(\vec
r_1)d^{b\dagger}_\downarrow(\vec
r_2)\right]u_\uparrow^{c\dagger}(\vec r_3)|0\rangle,
\end{eqnarray}
where $a$, $b$, and $c$ are color indices and the anti-commutation
relation of the non-relativistic quark creation and annihilation operators
is defined as $\{u^{a\dagger}_\alpha(\vec x), u^b_\beta(\vec
y)\}=\delta_{ab}\delta_{\alpha\beta}\delta^3(\vec x-\vec y) $ with $\alpha$ and
$\beta$ as spin indices. The spatial
part of the wave functions is common for all members of the baryon octet.
The SU(6) internal wave functions are listed in Table \ref{baryon} for easy reference.
\begin{table}[h]
\begin{tabular}{l}
\hline
$|p_\uparrow\rangle\sim\frac{1}{\sqrt{18}}\epsilon^{abc}[u^{a\dagger}_\downarrow
d_\uparrow^{b\dagger}-u^{a\dagger}_\uparrow
d^{b\dagger}_\downarrow]u_\uparrow^{c\dagger}|0\rangle;$\\
$|n_\uparrow\rangle\sim\frac{1}{\sqrt{18}}\epsilon^{abc}[d^{a\dagger}_\uparrow
u^{b\dagger}_\downarrow - d^{a\dagger}_\downarrow
u^{b\dagger}_\uparrow]d^{c\dagger}_\uparrow|0\rangle; $\\
$|\Lambda_\uparrow\rangle\sim\frac{1}{\sqrt{12}}
\epsilon^{abc}[u^{a\dagger}_\uparrow d^{b\dagger}_\downarrow -
u^{a\dagger}_\downarrow
d^{b\dagger}_\uparrow]s^{c\dagger}_\uparrow|0\rangle; $\\
$|\Sigma^+_\uparrow\rangle\sim\frac{1}{\sqrt{18}}\epsilon^{abc}[s^{a\dagger}_\downarrow
u^{b\dagger}_\uparrow - s^{a\dagger}_\uparrow
u^{b\dagger}_\downarrow]u^{c\dagger}_\uparrow|0\rangle;$\\
$|\Sigma^0_\uparrow\rangle\sim\frac{1}{6}\epsilon^{abc}[s^{a\dagger}_\uparrow
d^{b\dagger}_\downarrow u^{c\dagger}_\uparrow +
s^{a\dagger}_\uparrow d^{b\dagger}_\uparrow u^{c\dagger}_\downarrow
-2s^{a\dagger}_\downarrow d^{b\dagger}_\uparrow
u^{c\dagger}_\uparrow]|0\rangle;$\\
$|\Sigma^-_\uparrow\rangle\sim\frac{1}{\sqrt{18}}\epsilon^{abc}[s^{a\dagger}_\uparrow d^{b\dagger}_\downarrow - s^{a\dagger}_\downarrow d^{b\dagger}_\uparrow]d^{c\dagger}_\uparrow|0\rangle;$\\
$|\Xi^0_\uparrow\rangle\sim\frac{1}{\sqrt{18}}\epsilon^{abc}[s^{a\dagger}_\downarrow
u^{b\dagger}_\uparrow - s^{a\dagger}_\uparrow
u^{b\dagger}_\downarrow]s^{c\dagger}_\uparrow|0\rangle;$\\
$|\Xi^-_\uparrow\rangle\sim\frac{1}{\sqrt{18}}\epsilon^{abc}[s^{a\dagger}_\uparrow
d^{b\dagger}_\downarrow - s^{a\dagger}_\downarrow
d^{b\dagger}_\uparrow]s^{c\dagger}_\uparrow|0\rangle.$\\
\hline
\end{tabular}
\caption{SU(6) wave functions of baryon spin-1/2
octet.}\label{baryon}
\end{table}

Using Eqs. (\ref{3})-(\ref{8}), one can project operator $\bar u
i\gamma_5 u \bar d d$ into different irreducible representations of
the chiral group, $O_{11}^{ud,(\bar 3,3)}$, $O_{11}^{ud,(6,\bar6)}$,
and $O_{11}^{ud,(8,8)}$ as in Eq. (\ref{18}). Restricting to the
non-relativistic case, these operators become
\begin{eqnarray}\label{57}
O_{11}^{ud,(\bar3,3)}(x)&\simeq&-\frac{i}{8}:\left(u^{a\dagger}_\alpha(x)
u^a_\alpha(x) d^{b\dagger}_\beta(x)d^{b}_\beta(x) - d^{a\dagger}_\alpha(x) u^a_\alpha(x)
u^{b\dagger}_\beta(x) d^{b}_\beta(x)\right):\nonumber\\
O_{11}^{ud,(6,\bar6)}(x)&\simeq&-\frac{i}{8}:\left(u^{a\dagger}_\alpha(x)
u^a_\alpha(x) d^{b\dagger}_\beta(x)d^{b}_\beta(x) + d^{a\dagger}_\alpha(x) u^a_\alpha(x)
u^{b\dagger}_\beta(x) d^{b}_\beta(x)\right):\nonumber\\
O_{11}^{ud,(8,8)}(x)&\simeq&\frac{i}{4}:u^{a\dagger}_\alpha(x)u^a_\alpha(x)d^{b\dagger}_\beta(x)d^b_\beta(x):.
\end{eqnarray}
where $u$ and $d$ are non-relativistic two-component quark
annihilation operators, $a$ and $b$ label the color, $\alpha$ and
$\beta$ label the spin, and the ``: :'' means that the products of
the constituent quark fields are normal-ordered.

Considering the $(6,\bar6)$ component as an example, the simple
quark model gives the following matrix elements:
\begin{eqnarray}
\langle p_\uparrow(P)|O^{ud,(6,\bar6)}_{11}|p_\uparrow(
P)\rangle&=&\langle n_\uparrow(
P)|O^{ud,(6,\bar6)}_{11}|n_\uparrow(
P)\rangle\;=\;-\frac{i}{8}a \ , \nonumber\\
\langle \Sigma^+_\uparrow(
P)|O^{ud,(6,\bar6)}_{11}|\Sigma^+_\uparrow( P)\rangle&=&\langle
\Sigma^-_\uparrow(
P)|O^{ud,(6,\bar6)}_{11}|\Sigma^-_\uparrow(
P)\rangle\;=\;0 \ , \nonumber\\
\langle \Sigma^0_\uparrow(
P)|O^{ud,(6,\bar6)}_{11}|\Sigma^0_\uparrow(
P)\rangle&=&-\frac{i}{4}a \ , \nonumber\\
\langle \Lambda_\uparrow(
P)|O^{ud,(6,\bar6)}_{11}|\Lambda_\uparrow(
P)\rangle&=&0 \ , \nonumber\\
\langle \Xi^0_\uparrow(
P)|O^{ud,(6,\bar6)}_{11}|\Xi^0_\uparrow( P)\rangle&=&\langle
\Xi^-_\uparrow( P)|O^{ud,(6,\bar6)}_{11}|\Xi^-_\uparrow(
P)\rangle\;=\;0 \ ,
\end{eqnarray}
where $a=\int d^3r f^*(\vec P;\vec x,\vec x,\vec r)f(\vec P;\vec
x,\vec x,\vec r)$ is independent of $\vec x$. It is easy to check
that these matrix elements satisfy the symmetry conditions listed in
Table \ref{relations}. Using Eq. (\ref{25}), one can get the Wilson
coefficients for $(6,\bar6)$ hadronic operators defined in Eq.
(\ref{19'});
\begin{eqnarray}
C_6^{(1)}&=&-C_6^{(2)}\;=\;\frac{1}{8}\frac{\alpha^3}{(2\pi)^{3/2}};\nonumber\\
C_6^{(3)}&=&C_4^{(4)}\;=\;0 \ .
\end{eqnarray}
Expanding the hadronic operators to the first
order, one can get the P-odd, CP-odd three-point nucleon-pion
couplings,  $h_c$ and $h_n$. The result induced by $O^{ud,(6,\bar6)}_{11}$ is
\begin{equation}\label{60}
h_c=C^{ud}_{11}\alpha^3/({8\pi^{3/2}F_\pi})\simeq0.022C^{ud}_{11}\alpha^3/F_\pi\;,\;\;\;h_n=0  \ .
\end{equation}
In the same way one can calculate $h_c$ and $h_n$ induced by the
$(\bar3,3)$ and $(8,8)$ components of $\bar ui\gamma_5 u\bar d d$.
Taking into account the hermitian conjugate part of
each component, the contribution for $h_c$ and $h_n$ is doubled.

\subsubsection{MIT Bag Model}

The basic idea of the bag model is that valence quarks are confined
in a bag where the vacuum is in a phase different from the true QCD
vacuum. The inside has a constant energy-momentum density generating
a negative pressure, $B$, which is balanced by the positive pressure
of the quarks. The bag is usually taken as a sphere of radius $R_0$.
The quarks inside the bag move freely with the following wave
functions,
\begin{eqnarray}
\psi_{n,-1,1/2,m}(\vec r,t) &=&
\frac{N}{\sqrt{4\pi}}\left(\begin{array}{c}ij_0(\omega_{n,-1}r/R_0)\chi_m\\
-j_1(\omega_{n,-1}r/R_0)\vec\sigma\cdot\hat r\chi_m\\
\end{array}\right) \ .
\end{eqnarray}
The normalization factor of the above is
\begin{equation}
N(\omega_{n\kappa})=\left(\frac{\omega_{n\kappa}^3}{2R_0^3(\omega_{n\kappa}+\kappa)\sin^2\omega_{n\kappa}}\right)^{1/2} \ .
\end{equation}
The boundary condition gives the energy eigenvalue equation,
\begin{equation}\label{116}
\tan\omega_{n\kappa}=\frac{\omega_{n\kappa}}{\omega_{n\kappa}+\kappa}
\ ,
\end{equation}
and numerical calculation gives $\omega_0=2.043$.
The ground state of quarks is $\kappa=-1$, $n=0$ state. For the
baryon octet, all the quarks are in this state. Keeping only this,
the quark operator can be written as
\begin{equation}\label{66}
q(x)= \psi_{0,-1,1/2,m}(\vec
x)e^{-i\omega_{0,-1}t/R_0}b_{0,-1,1/2,m}+  ({\rm
anti\!\!-\!quark~creation})  \ .
\end{equation}
The physical meaning of the operator $b_m(0)$ is that it annihilates a
quark with quantum number described by the wave function
$\psi_{0,-1,1/2,m}$. Due to the assumption that inside the bag the
interaction between quarks and gluons is negligible, flavor and spin
automatically become good quantum numbers.

\begin{table}[h]
\begin{tabular}{|c|c|c|c|c|}
\hline  & \multicolumn{2}{c|}{ NR quark model}&\multicolumn{2}{c|}{MIT bag model}\\
\cline{2-5} Operators&$h_c/(\alpha^3/F_\pi)$&$h_n/(\alpha^3/F_\pi)$&$h_c/(1/(R_0^3F_\pi))$&$h_n/(1/(R_0^3F_\pi))$\\
\hline $\bar ui\gamma_5u\bar d d$&0.045&0.13&0.029&-0.024\\
\hline $\bar di\gamma_5d\bar u u$&0.045&-0.13&0.029&0.024\\
\hline $\bar ui\gamma_5u\bar s s$&0&0&0&0\\
\hline $\bar si\gamma_5s\bar u u$&0&0&0&0\\
\hline $\bar di\gamma_5d\bar s s$&0&0&0&0\\
\hline $\bar si\gamma_5s\bar d d$&0&0&0&0\\
\hline $\bar ui\gamma_5u\bar u u$&0.045&0&0.029&0\\
\hline $\bar di\gamma_5d\bar d d$&0.045&-0.13&0.029&-0.083\\
\hline $\bar si\gamma_5s\bar s s$&0&0&0&0\\
\hline\hline
       $\bar ui\gamma_5\sigma^{\mu\nu}u\bar d\sigma_{\mu\nu} d$&0.18&0&0.12&0\\
\hline $\bar di\gamma_5\sigma^{\mu\nu}d\bar u\sigma_{\mu\nu} u$&0.18&0&0.12&0\\
\hline $\bar ui\gamma_5\sigma^{\mu\nu}u\bar s\sigma_{\mu\nu} s$&0&0&0&0\\
\hline $\bar si\gamma_5\sigma^{\mu\nu}s\bar u\sigma_{\mu\nu} u$&0&0&0&0\\
\hline $\bar di\gamma_5\sigma^{\mu\nu}d\bar s\sigma_{\mu\nu} s$&0&0&0&0\\
\hline $\bar si\gamma_5\sigma^{\mu\nu}s\bar d\sigma_{\mu\nu} d$&0&0&0&0\\
\hline
\end{tabular}
\caption{P-odd, CP-odd three-point pion-nucleon vertices generated
by P-odd, CP-odd four-quark operators. The couplings induced by
operators constructed by two color-octet currents are equal to the
couplings induced by corresponding color-singlet operators
multiplying by $-2/3$. }\label{cpoddvertices}
\end{table}

We again take the $(6,\bar6)$ component of $\bar ui\gamma_5u \bar d
d$ as an example, which  can be written as
\begin{eqnarray}
O^{ud(6,\bar
6)}_{11}&\sim&-\frac{i}{2}\bar\psi_\lambda(\vec
x)\mathbb{P}_L\psi_\sigma(\vec x)\bar\psi_\rho(\vec
x)\mathbb{P}_L\psi_\tau(\vec x)\nonumber\\
&&\times\left[u^{a\dagger}_\lambda u^a_\sigma d^{b\dagger}_\rho
d^b_\tau + d^{a\dagger}_\lambda u^a_\sigma u^{b\dagger}_\rho
d^b_\tau\right] \ ,
\end{eqnarray}
where $a$ and $b$ are indices of color, $\lambda$, $\sigma$, $\rho$,
$\tau$ labeling the spin. The creation and annihilation operators
here are just like $b_{0,-1,1/2,m}$ in Eq. (\ref{66}). Using
\begin{eqnarray}
&&\bar\psi_\lambda(x)\mathbb{P}_L\psi_\sigma(x)\nonumber\\
&=&\frac{N^2}{8\pi}\left\{\left[j_0^2(\omega_0r/R_0)-j_1^2(\omega_0r/R_0)\right]\delta_{\lambda\sigma}-2ij_0(\omega
r/R_0)j_0(\omega r/R)\chi^\dagger_\lambda\vec\sigma\cdot\hat
r\chi_\sigma\right\} \ ,
\end{eqnarray}
and only keeping the terms which give non-vanishing contributions
after integrating over a spherical region, we have
\begin{eqnarray}
&&\bar\psi_\lambda(x)\mathbb{P}_L\psi_\sigma(x)\bar\psi_\rho(x)\mathbb{P}_L\psi_\tau(x)\nonumber\\
&\simeq&\frac{N^4}{64\pi^2}\left\{\left[j_0^2(\omega_or/R_0)-j_1^2(\omega_0r/R_0)\right]^2\delta_{\lambda\sigma}\delta_{\rho\tau}\right.\nonumber\\
&&\left.-4j_0^2(\omega_0r/R_0)j_1^2(\omega_0r/R_0)(\chi^\dagger_\lambda\vec\sigma\cdot\hat
r\chi_\sigma)(\chi^\dagger_\rho\vec\sigma\cdot\hat
r\chi_\tau)\right\}\ , \nonumber
\end{eqnarray}
where we neglect the term proportional to $\vec\sigma\cdot\hat r$.
In a proton state normalized to our convention before, the expectation value of
the operator can be written as
\begin{eqnarray}
&& \langle p_\uparrow|O^{ud(6,\bar6)}_{11}|p_\uparrow\rangle
=-\frac{i}{2}N(w_0)^4\frac{1}{64\pi^2}\int
d^3x\nonumber\\
&&\times
\left\{\left[j_0^2(\omega_or/R_0)-j_1^2(\omega_or/R_0)\right]^2\langle
p_\uparrow|A|p_\uparrow\rangle-4j_0^2(\omega_0r/R_0)j_1^2(\omega_0r/R_0)\langle
p_\uparrow|B|p_\uparrow\rangle\right\}\ ,
\end{eqnarray}
where
\begin{eqnarray}\label{71}
A&=&:u^{a\dagger}_\lambda u^a_\lambda d^{b\dagger}_\rho d^b_\rho: +
:d^{a\dagger}_\lambda u^a_\lambda u^{b\dagger}_\rho
d^b_\rho:\nonumber\\
B&=&:u^{a\dagger}_\lambda(\chi^\dagger_\lambda\vec\sigma\cdot\hat
r\chi_\sigma)u^a_\sigma
d^{b\dagger}_\rho(\chi^\dagger_\rho\vec\sigma\cdot\hat
r\chi_\tau)d^b_\tau:\nonumber\\
&&+:d^{a\dagger}(\chi^\dagger_\lambda\vec\sigma\cdot\hat
r\chi_\sigma)u^a_\sigma
u^{b\dagger}_\rho(\chi^\dagger_\rho\vec\sigma\cdot\hat
r\chi_\tau)d^b_\tau: \ .
\end{eqnarray}
A straightforward calculation gives
\begin{eqnarray}
\langle p_\uparrow|A|p_\uparrow\rangle&=&\langle n_\uparrow|A|n_\uparrow\rangle\;=\;1\nonumber\\
\langle \Sigma^+_\uparrow|A|\Sigma^+_\uparrow\rangle&=&\langle \Sigma^-_\uparrow|A|\Sigma^-_\uparrow\rangle\;=\;0\nonumber\\
\langle\Sigma^0_\uparrow|A|\Sigma^0_\uparrow\rangle&=&2\nonumber\\
\langle \Lambda_\uparrow|A|\Lambda_\uparrow\rangle&=&0\nonumber\\
\langle \Xi^0_\uparrow|A|\Xi^0_\uparrow\rangle&=&\langle
\Xi^-_\uparrow|A|\Xi^-_\uparrow\rangle\;=\;0 \ ,
\end{eqnarray}
and
\begin{eqnarray}
\langle p_\uparrow|B|p_\uparrow\rangle&=&\langle n_\uparrow|B|n_\uparrow\rangle\;=\;1/3\nonumber\\
\langle \Sigma^+_\uparrow|B|\Sigma^+_\uparrow\rangle&=&\langle \Sigma^-_\uparrow|B|\Sigma^-_\uparrow\rangle\;=\;0\nonumber\\
\langle\Sigma^0_\uparrow|B|\Sigma^0_\uparrow\rangle&=&2/3\nonumber\\
\langle \Lambda_\uparrow|B|\Lambda_\uparrow\rangle&=&0\nonumber\\
\langle \Xi^0_\uparrow|B|\Xi^0_\uparrow\rangle&=&\langle
\Xi^-_\uparrow|B|\Xi^-_\uparrow\rangle\;=\;0 \ .
\end{eqnarray}
Therefore we can get in the MIT bag model
\begin{eqnarray}
\langle p_\uparrow|O^{ud(6,\bar 6)}_{11}(x)|p_\uparrow\rangle&=&i\mathbb{A}+\frac{i}{3}\mathbb{B} \ ,
\end{eqnarray}
where
\begin{eqnarray}
\mathbb{A}&=&-\frac{1}{2}N(\omega_0)^4\frac{1}{16\pi}R_0^3\int_0^1\left(\frac{r}{R_0}
\right)^2d\left(\frac{r}{R_0}\right)\left[j_0^2(\omega_0r/R_0)-j_1^2(\omega_0r/R_0)\right]^2\ , \nonumber\\
\mathbb{B}&=&\frac{1}{2}N(\omega_0)^4\frac{1}{4\pi}R_0^3\int_0^1\left(\frac{r}{R_0}\right)^2d
\left(\frac{r}{R_0}\right)j_0^2(\omega_0r/R_0)j_1^2(\omega_0r/R_0) \ ,
\end{eqnarray}
and similarly for other matrix elements.

\begin{table}[h]
\begin{tabular}{|c|c|c|c|c|}
\hline  & \multicolumn{2}{c|}{ NR quark model}&\multicolumn{2}{c|}{MIT bag model}\\
\cline{2-5} Operators&$h_c/(C_4\alpha^3/F_\pi)$&$h_n/(C_4\alpha^3/F_\pi)$&$h_c/(C_4/(R_0^3F_\pi))$&$h_n/(C_4/(R_0^3F_\pi))$\\
\hline $\bar ui\gamma_5u\bar d d$&$0.0883$&$0.374$&$0.0560$&$-0.0690$\\
\hline $\bar di\gamma_5d\bar u u$&$0.0883$&$-0.374$&$0.0560$&$0.0690$\\
\hline $\bar ui\gamma_5 t^a u\bar d t^a d$&$-0.0343$&$-0.0759$&$-0.0222$&$0.0140$\\
\hline $\bar di\gamma_5 t^a d\bar u t^a u$&$-0.0343$&$0.0759$&$-0.0222$&$-0.0140$\\
\hline $\bar ui\gamma_5u\bar u u$&$0.0883$&$0$&$0.0569$&$0$\\
\hline $\bar di\gamma_5d\bar d d$&$0.0883$&$-0.255$&$0.0569$&$-0.163$\\
\hline $\bar ui\gamma_5 t^a u\bar u t^a u$&$-0.0343$&$0$&$-0.0221$&$0$\\
\hline $\bar di\gamma_5 t^a d\bar d t^a d$&$-0.0343$&$0.0991$&$-0.0221$&$0.0633$\\
\hline\hline
       $\bar ui\gamma_5\sigma^{\mu\nu}u\bar d\sigma_{\mu\nu} d$&$-0.0397$&$0$&$-0.0230$&$0$\\
\hline $\bar ui\gamma_5\sigma^{\mu\nu} t^a u\bar d\sigma_{\mu\nu} t^a d$&$-0.268$&$0$&$-0.180$&$0$\\
\hline
\end{tabular}
\caption{Same as Table \ref{cpoddvertices}, except the matrix
elements are quoted here at the scale $\mu=4\pi F_\pi$ assuming the
quark model scale of 400 MeV.}\label{cpoddvertices1}
\end{table}

Then, using the method we used in the non-relativistic quark model,
we can get $h_c$ and $h_n$ induced by $O^{ud(6,\bar6)}_{11}$,
\begin{equation}
h_c=\frac{0.015C^{ud}_{11}}{R^3_0 F_\pi}\;,\;\;\;h_n=0 \ .
\end{equation}
One can compare this with the result from the non-relativistic quark
model in Eq. (\ref{60}), where $h_c$ is proportional to $\alpha^3$.
From the definition of $\psi$ below Eq. (\ref{55}), $1/\alpha$ can
also be seen as the radius of the baryon. It is well known that
$1/\alpha=0.5 $ fm gives a too small value for the proton's charge
radius and the pion cloud is usually invoked to gap it. On the other
hand, the bag radius is usually taken to be 1.0 fm, which will give
a considerably smaller $h_c$. In any case, it is reasonable to
consider $R_0\sim 1/\alpha$ and take the non-relativistic quark
model result as the representative.

The couplings $h_c$ and $h_n$ induced by color-singlet four-quark
operators are listed in Table \ref{cpoddvertices} and those by
color-octet operators are equal to the above multiplying by $-2/3$.
In Table \ref{cpoddvertices}, many four-quark operators yield zero
$h_c$ and $h_n$ because we neglect the ``sea quark'' contribution.
By making the four-quark operators normal ordered in Eq. (\ref{57})
and (\ref{71}), one cannot get any contribution to $h_c$ and $h_n$
from four-quark operators containing strange quarks.

Model calculations do not have explicit QCD scale dependence. To
match the results with QCD matrix elements, we have to assume a
model scale and using perturbative QCD (pQCD) evolution to run them
to appropriate perturbative scale, for which we choose to be $\mu=
4\pi F_\pi$.  In this work, we assume the model scale to be at 400
MeV and $\Lambda_{\rm QCD}=250$ MeV and take into account the pQCD
effect using one-loop renormalization group equation to run the
operators down to the energy scale of the model. At this low energy
regime the strong coupling is large and the one-loop pQCD evolution
is by no means accurate, but it may still serve as an estimate of
the pQCD effect. The matrix elements at scale $\mu$ are shown in
Table \ref{cpoddvertices1}.

\subsubsection{Contribution from odd-parity resonances}

\begin{figure}[hbt]
\begin{center}
\includegraphics[width=6cm]{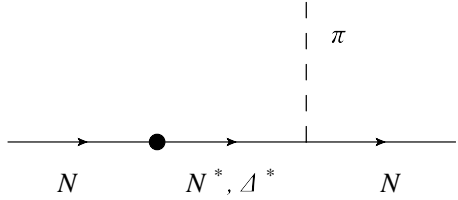}
\caption{P-odd and CP-odd pion-nucleon coupling generated by the
four-quark operators through parity-odd resonances, where the black
dot is the CP-odd, four-quark operator, $N^*$ and $\Delta^*$ are the
CP-odd excited states. }\label{treepion}
\end{center}
\end{figure}

The P-odd and CP-odd quark operators can also generate a CP-odd
pion-nucleon interaction through the parity-odd excited resonances
which is shown in Fig. \ref{treepion}. The P-odd and CP-odd quark
operators can generate mixings between nucleons and parity-odd
excited resonances which can be calculated using quark
models~\cite{Ecker:1983dj}. Take the operator $O^{(ud)}_{11}=\bar
ui\gamma_5u\bar d d$ and the intermediate state N(1535) as an
example, using the harmonic oscillator non-relativistic quark model
the mass mixing between neutron and $N(1535)$ resonance can be
estimated as $\delta = m_c \omega^2 / (8\sqrt{3}\pi^{3/2})$, where
$m_c\approx\omega\approx300$ MeV are the constituent quark mass and
the frequency of the harmonic oscillator, respectively. The
resonance can decay into a nucleon plus a pion, the partial decay
width is about 50 MeV~\cite{Amsler:2008zzb}. The effective
Lagrangian for this process can be written as
\begin{equation}
{\cal L}_{N^*} = g_{N^*} \bar N N^* \pi +h.c.\ ,
\end{equation}
where as an order-of-magnitude estimate we discard the isospin
quantum number. Then, from the partial decay width one can get
$g_{N^*}\sim{\cal O}(1)$. The P-odd and CP-odd pion-nucleon coupling
induced by this mixing can be written as
\begin{equation}
h_{mix} = \frac{C_4 g_{N^*}\delta}{M_{N^*}-m_n}\approx
6\times10^{-4} C_4 {\rm GeV}^2\ ,
\end{equation}
where $C_4$ is the Wilson coefficient of the four-quark operator.
Compared with the direct matching contribution listed in Table
\ref{cpoddvertices}, one can see that $h_{mix}$ is about two orders
of magnitude smaller and therefore its contribution to nEDM is
negligible.

The contribution from Fig. \ref{treepion} can be seen as a one-loop
contribution since the intermediate resonances may also be described
as scattering states of pion and nucleon. Therefore, this
contribution is suppressed by a loop factor.

\subsection{Tree-Level CP-Odd Mass of Neutron}

The nucleon CP-odd observables receive contributions from its CP-odd mass term
$m'\bar \psi i\gamma_5 \psi$. In $\chi$PT, there are also two sources of CP-odd mass:
that induced by the condensates of meson fields, namely $\langle\pi^0\rangle$ and
$\langle\eta\rangle$, and that from the direct matching contribution of
the four-quark operators.

\subsubsection{Meson Condensates}

The relevant terms contributing to the CP-odd mass of neutron can be
read from expanding Eq. (\ref{77}), which gives
\begin{eqnarray}
&&\bar n i\gamma_5 n
\frac{1}{F_\pi}\left\{-d_1[(m_u-m_d)\langle\pi^0\rangle+\frac{1}{\sqrt{3}}(m_u+m_d)
\langle\eta\rangle-\frac{2}{\sqrt{3}}m_s\langle\eta\rangle]\right.\nonumber\\
\;\;\;\;&&\left.+d_2\frac{2}{\sqrt{3}}m_s\langle\eta\rangle + d_3
m_d(\langle\pi^0\rangle-\frac{1}{\sqrt{3}}\langle\eta\rangle)\right\}\;,
\end{eqnarray}
where $d_1$, $d_2$ and $d_3$ can be related to the discrepancy of
the Goldberger-Treiman relation, and the values $d_2$ and $d_3$ have
been determined in the literature~\cite{Goity:1999by}.
\begin{equation}
d_2 = -2B_0 m_0 (D_{19} - F_{19})\;,\;\; d_3 = -2 B_0 m_0 (D_{19} +
F_{19})\;,
\end{equation}
where $m_0$ is the common octet mass in the chiral limit, and
\begin{equation}
m_0 F_{19}\approx -0.2\;, \;\;m_0 D_{19}\approx -0.4\;.
\end{equation}
Note that the signs of the $F_{19}$ and $D_{19}$ here are different
from those in Ref. \cite{Goity:1999by}. Since $d_1$ has not been
determined from isospin-violation effect, we will set it to be zero
in the following calculation. One should note that disregarding
$d_1$ leads to some errors because $m_s\langle\eta\rangle$ might be
the same order as $m_d\langle\pi^0\rangle$.

\begin{table}[h]
\begin{tabular}{|c|c|c|c|}

\hline Operators & $m'_n/(10^{-3}C_4 B_0^2\;{\rm GeV})$&Operators & $m'_n/(10^{-3}C_4B_0^2\;{\rm GeV})$\\
\hline $\bar u i\gamma_5 u \bar d d$&$-8.8$ &$\bar u t^a i\gamma_5 u \bar d t^a d$&$0$\\
\hline $\bar d i\gamma_5 d \bar u u$&$5.7$&$\bar d t^a i\gamma_5 d \bar u t^a u$&$0$\\
\hline $\bar u i\gamma_5 d \bar s s$&$-8.8$&$\bar u t^a i\gamma_5 u \bar s t^a s$&$0$\\
\hline $\bar s i\gamma_5 s \bar u u$&$3.2$&$\bar s t^a i\gamma_5 s \bar u t^a u$&$0$\\
\hline $\bar d i\gamma_5 d \bar s s$&$5.7$&$\bar d t^a i\gamma_5 d \bar s t^a s$&$0$\\
\hline $\bar s i\gamma_5 s \bar d d$&$3.2$&$\bar s t^a i\gamma_5 s \bar d t^a d$&$0$\\
\hline $\bar u i\gamma_5 u \bar u u$&$-7.4$&$\bar u t^a i\gamma_5 u \bar u t^a u$&$2.0$\\
\hline $\bar d i\gamma_5 d \bar d d$&$4.7$&$\bar d t^a i\gamma_5 d \bar d t^a d$&$-1.3$\\
\hline $\bar s i\gamma_5 s \bar s s$&$2.6$&$\bar s t^a i\gamma_5 s \bar s t^a s$&$0.7$\\
\hline
\end{tabular}\caption{CP-odd mass of the neutron induced by meson
condensates. Contributions from operators made of tensor currents are neglected due to
the large-$N_C$
suppression.}\label{massvev}
\end{table}

\subsubsection{Direct Contribution}

The leading-order expansion of the tilded hadronic operators listed
in Table \ref{Hadronic} are hermitian.  Take $\tilde O_6^{(2)}$ as
an example. It can be written as
\begin{equation}\label{82}
\tilde O_6^{(2)}\simeq \bar p i\gamma_5 p + \bar n i\gamma_5 n +
\frac{1}{3}\bar\Lambda i\gamma_5\Lambda + \bar \Sigma^0
i\gamma_5\Sigma^0 + \bar \Sigma^+ i\gamma_5\Sigma^+ + \bar \Sigma^-
i\gamma_5\Sigma^-\;,
\end{equation}
which gives a CP-odd mass of neutron. To calculate the matching
coefficients, we can see from above that the leading-order expansion
is parity-odd, and we need to calculate a parity-odd quantity. The
simplest is $\Delta\vec s\cdot\Delta\vec p$, where $\Delta \vec s$
is the spin difference between the initial and final states and
$\Delta \vec p$ is the momentum difference between the initial and
final states.

\begin{table}[h]
\begin{tabular}{|c|c|c|c|}
\hline Operators & CP-odd mass$/(\alpha^3C_4)$&Operators &
CP-odd mass $/(\alpha^3C_4)$\\
\hline $\bar ui\gamma_5 u\bar d d$&$0.0635$&$\bar ui\gamma_5\sigma^{\mu\nu} u\bar d\sigma_{\mu\nu} d$ &$-0.127$ \\
\hline $\bar di\gamma_5 d\bar u u$&$-0.127$&$-$ &$-$ \\
\hline $\bar ui\gamma_5 u\bar s s$&0&$\bar ui\gamma_5\sigma^{\mu\nu} u\bar s\sigma_{\mu\nu} s$&0\\
\hline $\bar si\gamma_5 s\bar u u$&0&$-$&$-$\\
\hline $\bar di\gamma_5 d\bar s s$&0&$\bar di\gamma_5\sigma^{\mu\nu} d\bar s\sigma_{\mu\nu} s$&0\\
\hline $\bar si\gamma_5 s\bar d d$&0&$-$&$-$\\
\hline $\bar ui\gamma_5 u\bar u u$&0&$-$&$-$\\
\hline $\bar di\gamma_5 d\bar d d$&$-0.127$&$-$&$-$\\
\hline $\bar si\gamma_5 s\bar s s$&0&$-$&$-$\\
\hline
\end{tabular}\caption{CP-odd mass of neutron induced directly by color-singlet four-quark operators.
The CP-odd mass induced by color-octet four-quark operators are equal
to the one induced by corresponding color-singlet operators
multiplied by $-2/3$.}\label{massdi}
\end{table}

\begin{table}[h]
\begin{tabular}{|c|c|c|c|}
\hline Operators & CP-odd mass$/(\alpha^3C_4)$&Operators &
CP-odd mass $/(\alpha^3C_4)$\\
\hline $\bar ui\gamma_5 u\bar d d$&$0.212$&$\bar ui\gamma_5\sigma^{\mu\nu} u\bar d\sigma_{\mu\nu} d$ &$0.0280$ \\
\hline $\bar di\gamma_5 d\bar u u$&$-0.336$&$\bar ui\gamma_5\sigma^{\mu\nu} t^a u\bar d\sigma_{\mu\nu} t^a d$ &$0.189$ \\
\hline $\bar ui\gamma_5 t^a u\bar d t^a d$&$-0.0314$&$-$ &$-$ \\
\hline $\bar di\gamma_5 t^a d\bar u t^a u$&$0.0799$&$-$ &$-$ \\
\hline $\bar ui\gamma_5 u\bar u u$&0&$\bar ui\gamma_5 t^a u\bar u t^a u$&$0$\\
\hline $\bar di\gamma_5 d\bar d d$&$-0.249$&$\bar di\gamma_5 t^a d\bar d t^a d$&$0.0968$\\
\hline
\end{tabular}\caption{Same as Fig. \ref{massdi}. The matrix elements are now
evolved to the scale where $\mu = 4\pi F_\pi$. }\label{massdi1}
\end{table}

In the non-relativistic quark model, take the $(6,\bar6)$ components
of $\bar u i\gamma_5 u\bar d d$, as an example, to calculate the
matrix elements proportional to $\Delta\vec s\cdot\Delta\vec p$; the
relevant part of the four-quark operator can be written as
\begin{eqnarray}
O^{ud,(6,\bar6)}_{11}&\sim&-\frac{i}{8}\left\{\frac{i}{2m_C}:[\nabla\cdot(u^\dagger\vec\sigma
u)](d^\dagger d): + \frac{i}{2m_C}:(u^\dagger
u)[\nabla\cdot(d^\dagger\vec\sigma d)]:\right.\nonumber\\
&&\left.-\frac{i}{2m_C}:[\nabla\cdot(d^\dagger\vec\sigma
u)](u^\dagger d):-\frac{i}{2m_C}:(d^\dagger
u)[\nabla\cdot(u^\dagger\vec\sigma d)]:\right\}\;,
\end{eqnarray}
where $u$ and $d$ are two-component quark operators, $m_C$ is the
mass of the constituent quark which is set to be one-third of the
nucleon mass. The wave functions of baryons in the non-relativistic
quark model are listed in Eq. (\ref{95}) and Table \ref{baryon}.
Then using the same method as described in the last section one can
get the CP-odd mass of the neutron directly induced by the tilded
operators, and the results are listed in Table \ref{massdi}. After
the leading-order QCD evolution to the scale where $\mu = 4\pi
F_\pi$, the result is shown in Table \ref{massdi1}.

\subsubsection{Contribution to CP-Odd Meson-Nucleon Coupling}

If rotating away the CP-odd nucleon mass through $U_A(1)$ transformation,
one can generate new contributions to the CP-odd meson-nucleon coupling from CP-even chiral operators.
However, this contribution is of higher order in chiral power counting because
all the CP-even meson-nucleon interactions are suppressed in the chiral limit,
whereas the CP-odd coupling we considered in the previous subsections
are not.

\section{Four-Quark Contribution to NEUTRON EDM IN $\chi$PT}

In this section, we study the CP-odd four-quark contributions to the
neutron EDM in $\chi$PT. The approach here is completely general and is
applicable to any CP-odd quark-gluon operators. Some results presented can
be found in the literature; however, to our knowledge, this is the most systematic and
thorough discussion in the context of the CP-odd four-quark operators.
In the last subsection, we make a comparison of the four-quark contributions in
different approximations of non-perturbative QCD physics.

In $\chi$PT, the leading contributions come from many different
sources. Since the CP-violating pion-nucleon couplings are ${\cal
O}(1)$, the pion loop contribution to the neutron EDM is ${\cal
O}(1)$, apart from possible enhancement by chiral logarithms. On the
other hand, the direct matching contribution is also ${\cal O}(1)$,
along with the pion condensate contribution through photo-production
amplitudes. Finally, the CP-odd mass terms contribute through the
nucleon magnetic moment after chiral rotation. This contribution is
again ${\cal O}(1)$ in chiral power counting. We will consider all
these leading contributions in the following subsections. We ignore
the subleading contribution in this work.

\subsection{Direct Matching from Quark Model}

We have first considered the direct matching contribution from the
four-quark operators to the neutron EDM in Sec. IV. When any CP-odd
quark-gluon operator is matched in $\chi$PT, there appear many
tree-level neutron EDM-like operators in the chiral
Lagrangian~\cite{Pich:1991fq}. We do not have much to say about the
size of the Wilson coefficients other than they are ${\cal O}(1)$ in
chiral power counting. Since they also serve as the counter terms
for ultraviolet-divergent chiral-loop calculations, they depend on
the regularization scheme and subtraction scale. In this work, we
choose to estimate this contribution using nucleon models with
dipole excitations into odd-parity resonances, such as $S_{11}$,
following the work in \cite{Ecker:1983dj}.

\begin{figure}[hbt]
\begin{center}
\includegraphics[width=6cm]{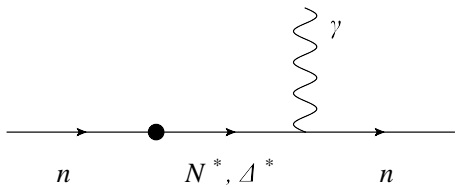}
\caption{Direct calculation of the neutron EDM in quark models. The
neutron makes a transition to a CP-odd excited state and goes back
via electromagnetic interaction, where the black dot is the CP-odd,
four-quark operator, $N^*$ and $\Delta^*$ are the CP-odd excited
states. }\label{tree}
\end{center}
\end{figure}

We use the non-relativistic quark model with harmonic oscillator
potentials to estimate the contribution from the first CP-odd excited
states, which is shown in Fig. \ref{tree}. The wave functions of the lowest
CP-odd excited states can be written as
\begin{eqnarray}
|N^*_\uparrow\rangle &=& N_1 \epsilon^{abc}\int d^3r_1 d^3r_2 d^3r_3
\exp\left(\frac{i\vec P\cdot \vec
R}{\sqrt{3}}-\frac{\alpha^2}{2}(\rho^2+\lambda^2)\right)\nonumber\\
&&\left\{(\lambda_x+i\lambda_y)[u^{a\dagger}_\downarrow(r_1)d^{b\dagger}_\uparrow(r_2)d^{c\dagger}_\downarrow(r_3)-u^{a\dagger}_\uparrow(r_1)d^{b\dagger}_\downarrow(r_2)d^{c\dagger}_\downarrow(r_3)]|0\rangle\right.\nonumber\\
&&\left.-\lambda_z[u^{a\dagger}_\uparrow(r_1)d^{b\dagger}_\downarrow(r_2)d^{c\dagger}_\uparrow(r_3)-u^{a\dagger}_\downarrow(r_1)d^{b\dagger}_\uparrow(r_2)d^{c\dagger}_\uparrow(r_3)]|0\rangle\right\}\;;\nonumber\\
|\Delta^*_\uparrow\rangle &=& N_2 \epsilon^{abc}\int d^3r_1 d^3r_2
d^3r_3 \exp\left(\frac{i\vec P\cdot \vec
R}{\sqrt{3}}-\frac{\alpha^2}{2}(\rho^2+\lambda^2)\right)\nonumber\\
&&\left\{(\lambda_x+i\lambda_y)[2u^{a\dagger}_\downarrow(r_1)d^{b\dagger}_\downarrow(r_2)d^{c\dagger}_\uparrow(r_3)+d^{a\dagger}_\downarrow(r_1)d^{b\dagger}_\downarrow(r_2)u^{c\dagger}_\uparrow(r_3)]|0\rangle\right.\nonumber\\
&&\left.-\lambda_z[2u^{a\dagger}_\uparrow(r_1)d^{b\dagger}_\uparrow(r_2)d^{c\dagger}_\downarrow(r_3)+d^{a\dagger}_\uparrow(r_1)d^{b\dagger}_\uparrow(r_2)u^{c\dagger}_\downarrow(r_3)]|0\rangle\right\}\;.
\end{eqnarray}
In the above formulas $\lambda_x$, $\lambda_y$ and $\lambda_z$ are
the $x$, $y$ and $z$ components of $\lambda$, respectively. $N_1$
and $N_2$ are normalization factors of the states with $
N_1={2^{1/2}\alpha^4}/({3^{9/4}\pi^{3/2}})\;,\;\;N_2={\alpha^4}/({2^{1/2}3^{9/4}\pi^{3/2}})\;.$

\begin{table}[h]
\begin{tabular}{|c|c|c|c|}
\hline Operators & nEDM$/(e\alpha C_4)$&Operators &
nEDM$/(e\alpha C_4)$\\
\hline $\bar ui\gamma_5 u\bar d d$&$-\frac{1}{6\sqrt{2}\pi^{3/2}}$&$\bar ui\gamma_5 t^a u\bar d t^a d$ &$\frac{1}{9\sqrt{2}\pi^{3/2}}$ \\
\hline $\bar di\gamma_5 d\bar u u$&$-\frac{1}{3\sqrt{2}\pi^{3/2}}$&$\bar di\gamma_5 t^a d\bar u t^a u$ &$\frac{\sqrt{2}}{9\pi^{3/2}}$ \\
\hline $\bar ui\gamma_5\sigma^{\mu\nu} u\bar d\sigma_{\mu\nu} d$&$\frac{1}{\sqrt{2}\pi^{3/2}}$&$\bar ui\gamma_5\sigma^{\mu\nu} t^a u\bar d\sigma_{\mu\nu} t^a d$ &$-\frac{\sqrt{2}}{3\pi^{3/2}}$ \\
\hline $\bar ui\gamma_5 u\bar u u$&$0$&$\bar ui\gamma_5 t^a u\bar u t^a u$ &$0$ \\
\hline $\bar di\gamma_5 d\bar d d$&$0$&$\bar di\gamma_5 t^a d\bar d t^a d$ &$0$ \\
\hline
\end{tabular}\caption{nEDM contributed from first excited CP-odd states in the non-relativistic quark model, where
$C_4$ is the Wilson coefficients of the quark models, $\alpha$ is
defined below Eq. (\ref{55}). The unit of nEDM used here is
$e\cdot$GeV$^{-1}$, which is different from the traditional one
$e\cdot$cm due to that the Wilson coefficients of the four-quark
operators are unknown which are always in the unit of GeV$^{-2}$.
The translation between the two units is $e\cdot {\rm
GeV}^{-1}\simeq2\times 10^{-14}e{\rm \cdot cm}$.}\label{counter}
\end{table}
\begin{table}[h]
\begin{tabular}{|c|c|c|c|}
\hline Operators & nEDM$/(10^{-3} e C_4 \rm{GeV})$&Operators &
nEDM$/(10^{-3} e C_4 \rm{GeV})$\\
\hline $\bar ui\gamma_5 u\bar d d$&$-37.6$&$\bar ui\gamma_5 t^a u\bar d t^a d$ &$3.80$ \\
\hline $\bar di\gamma_5 d\bar u u$&$-62.6$&$\bar di\gamma_5 t^a d\bar u t^a u$ &$8.87$ \\
\hline $\bar ui\gamma_5\sigma^{\mu\nu} u\bar d\sigma_{\mu\nu} d$&$77.5$&$\bar ui\gamma_5\sigma^{\mu\nu} t^a u\bar d\sigma_{\mu\nu} t^a d$ &$-103$ \\
\hline $\bar ui\gamma_5 u\bar u u$&$0$&$\bar ui\gamma_5 t^a u\bar u t^a u$ &$0$ \\
\hline $\bar di\gamma_5 d\bar d d$&$0$&$\bar di\gamma_5 t^a d\bar d t^a d$ &$0$ \\
\hline
\end{tabular}\caption{Same as Table XIII, except the renormalization scale is now at $4\pi F_\pi$.}\label{counterafterrunning}
\end{table}

The results are shown in Table \ref{counter}, which agree with the
results extracted from Ref. \cite{Ecker:1983dj}. We also need to
take into account the evolution of the operators between $4\pi
F_\pi$ and the energy scale of the quark model. The results are
shown in Table \ref{counterafterrunning} with $\alpha=0.41$ GeV.

\subsection{Meson Condensate Contribution through Photo-Pion Production}

\begin{figure}[hbt]
\begin{center}
\includegraphics[width=4cm]{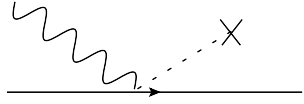}
\caption{Pion-photoproduction diagram with the pion field
annihilated by the four-quark operator into the vacuum, where the cross
is a four-quark operator. }\label{photoproduction}
\end{center}
\end{figure}

In photon-pion production, there are CP-even electric-dipole
couplings between the baryon-octet and electromagnetic fields
through using $f^{\mu\nu}_\pm$~\cite{Pich:1991fq}.  Some of these
couplings can generate the neutron EDM if they violate the chiral
symmetry through the quark masses and at the same time the meson
fields acquire vacuum condensates through the CP-odd four-quark
operators. In more physical language, the contact terms for the
pion-photoproduction processes give rise to the neutron EDM through
the diagram in Fig. \ref{photoproduction}.
Although the electromagnetic
field also violates chiral symmetry,
it cannot generate an EDM through meson condensates by itself---a
quark mass factor is essential.

\begin{table}[h]
\begin{tabular}{|c|c|c|c|}

\hline Operators & $d_{\pi\gamma}/(10^{-3}eC_4 B_0^2\;{\rm GeV}^{-1})$&Operators & $d_{\pi\gamma}/(10^{-3}eC_4B_0^2\;{\rm GeV}^{-1})$\\
\hline $\bar u i\gamma_5 u \bar d d$&$10.6$ &$\bar u t^a i\gamma_5 u \bar d t^a d$&$0$\\
\hline $\bar d i\gamma_5 d \bar u u$&$-10.5$&$\bar d t^a i\gamma_5 d \bar u t^a u$&$0$\\
\hline $\bar u i\gamma_5 d \bar s s$&$10.6$&$\bar u t^a i\gamma_5 u \bar s t^a s$&$0$\\
\hline $\bar s i\gamma_5 s \bar u u$&$-0.12$&$\bar s t^a i\gamma_5 s \bar u t^a u$&$0$\\
\hline $\bar d i\gamma_5 d \bar s s$&$-10.5$&$\bar d t^a i\gamma_5 d \bar s t^a s$&$0$\\
\hline $\bar s i\gamma_5 s \bar d d$&$-0.12$&$\bar s t^a i\gamma_5 s \bar d t^a d$&$0$\\
\hline $\bar u i\gamma_5 u \bar u u$&$8.87$&$\bar u t^a i\gamma_5 u \bar u t^a u$&$-2.37$\\
\hline $\bar d i\gamma_5 d \bar d d$&$-8.77$&$\bar d t^a i\gamma_5 d \bar d t^a d$&$2.34$\\
\hline $\bar s i\gamma_5 s \bar s s$&$-0.10$&$\bar s t^a i\gamma_5 s \bar s t^a s$&$0.03$\\
\hline
\end{tabular}\caption{nEDM induced by meson
condensates through pion-photoproduction. Contribution from
operators constructed by tensor operators are neglected due to the
large-$N_C$ suppression.}\label{pionphoto}
\end{table}

The terms of interest are made of linear products of baryon fields
$\bar B$ and $B$, $\chi_-$ and $f_+$~\cite{Pich:1991fq},
\begin{equation}
{\cal L}_{\pi\gamma}^C = \frac{1}{16\pi^2F_\pi^2} \left[\delta_1{\rm
Tr}[\bar B\sigma_{\mu\nu}\gamma_5\{\chi_-,f^{\mu\nu}_+\}B] +
\delta_2{\rm Tr}[\bar B\sigma_{\mu\nu}\gamma_5 f^{\mu\nu}_+B]{\rm
Tr}[\chi_-] + ... \right]\ ,
\end{equation}
where we have shown two of the ten possible terms. It is difficult,
however, to extract the Wilson coefficients $\delta_i$ directly from
experimental data. Some of the coefficients have been estimated by
calculating the contribution from the excited baryon states in the
context of the two-flavor scenario~\cite{Bernard:1994gm}. In the
two-flavor scenario, neglecting the isospin violation generated by
the difference between the up and down quark masses, the terms
relevant to nEDM can be written as
\begin{equation}
{\cal L}_{\pi\gamma}^{2-flavor} = \bar
N\gamma_5\sigma_{\mu\nu}[(a_1^p-a_1^n)f^{\mu\nu}_+ + a_1^n {\rm
Tr}(f^{\mu\nu}_+)]\chi_-N\;,
\end{equation}
where $N = \left(\begin{array}{c}p\\n\\ \end{array}\right)$, and in
the two-flavor case, $f^{\mu\nu}_+\equiv e(\xi^\dagger Q\xi + \xi
Q\xi^\dagger)F^{\mu\nu}$, in which $Q=(1+\tau^3)/2$. Expanding
$f_+^{\mu\nu}$ and $\chi_-$, we can get the nEDM induced by the
condensate of $\pi^0$;
\begin{equation}
d_{\pi\gamma} = -\frac{8e a_1^n B_0 (m_u+m_d)
\langle\pi^0\rangle}{F_\pi}\;.
\end{equation}
From Ref. \cite{Bernard:1994gm}, one can get the contribution to
$a_1$ from $\Delta$ and $\rho$ internal states, which is
\begin{equation}
a_1 = -0.156 {\rm GeV}^{-3}\;.
\end{equation}
Using this, one can estimate the nEDM induced by the pion condensate,
as shown in Table \ref{pionphoto}.

\subsection{CP-Odd Baryon Mass Contribution}

\begin{figure}[hbt]
\begin{center}
\includegraphics[width=5cm]{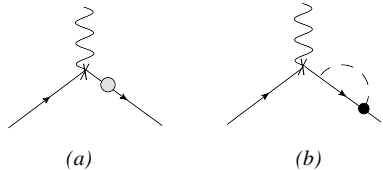}
\caption{The CP-odd mass of neutron turns the tree level magnetic
moment into an EDM. The cross is the tree level magnetic moment, the
gray dot is the CP-odd mass of the neutron and the black dot is the
CP-odd pion-nucleon coupling.}\label{imaginarymass}
\end{center}
\end{figure}

The CP-odd baryon-mass terms considered in the previous section
generate a CP-odd part of the baryon wave function. This part can transform a magnetic moment
term into an EDM contribution. The physics of this is shown in Fig. \ref{imaginarymass}.

The mass terms of the neutron can be written as
\begin{equation}
{\cal L}_{mass}=-m_n\bar n n - m'_n \bar n i\gamma_5 n \ .
\end{equation}
Note that the neutron field $n$ here is already redefined using the
transformation in Eq. (\ref{cotransformation})
after taking into account the meson condensate effect as discussed
in the previous section. Redefining the neutron field again through
a chiral rotation,
\begin{equation}\label{redef}
n=\exp(-i\frac{m_n'}{2m_n})\gamma_5 n' \ ,
\end{equation}
the mass term becomes the standard one,
\begin{equation}
{\cal L}_{\rm mass}=-m_n \bar n' n' \ .
\end{equation}
On the other hand, the tree level anomalous magnetic moment of the neutron can be written
as
\begin{equation}
{\cal L}_{\rm mag. mom.} = -\frac{1}{4}\frac{\kappa_n}{m_n}\bar n\sigma^{\mu\nu}n F_{\mu\nu} \
.
\end{equation}
The redefinition in Eq. (\ref{redef}) generates a neutron
EDM,
\begin{equation}
d^{\rm EDM}_{\rm CP-odd mass}=-\frac{\kappa_n m_n'}{2 m_n^2} \ .
\end{equation}
The experimental values of the anomalous magnetic dipole moments of
the nucleons are $ \kappa_p=1.7928, \;\kappa_n=-1.9131 . $ The
numerical values of this contribution have been shown in Tables
\ref{singletedm} and \ref{octetedm}. For the tensor operator $\bar
ui\gamma_5\sigma^{\mu\nu} u \bar d \sigma_{\mu\nu}d$, the
contribution from the CP-odd mass of the nucleon is particularly
large. The CP-odd mass also gets a quantum correction shown in
diagram (b) of Fig. \ref{imaginarymass}. It is easy to see that this
term does not have any chiral enhancement and is of a higher-order
effect.

\subsection{Leading Chiral Loop Contribution}

\begin{figure}[hbt]
\begin{center}
\includegraphics[width=6cm]{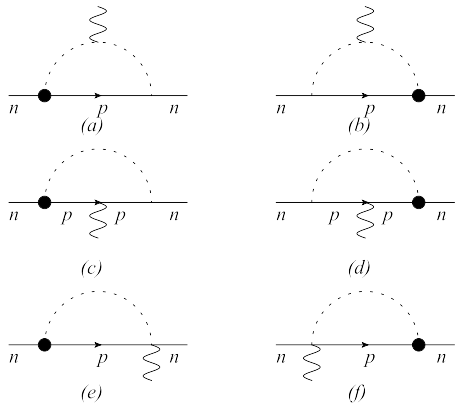}
\caption{Charged-pion loop contribution to neutron EDM (without the anomalous magnetic moment), where
the black dots represent the CP-odd vertices. }\label{pionloop}
\end{center}
\end{figure}

\begin{figure}[hbt]
\begin{center}
\includegraphics[width=6cm]{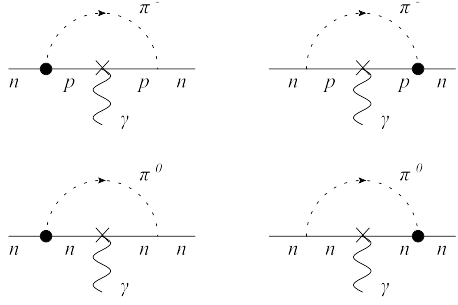}
\caption{Contribution from the tree level anomalous magnetic moments
of proton and neutron, where the crosses are anomalous magnetic
moments of nucleons and the dots are CP-odd vertices. }\label{amdm}
\end{center}
\end{figure}

The contribution we have considered so far has a smooth chiral
limit, i.e., regular as the quark masses go to zero. The leading
contribution in the chiral limit, however, involves the pion loop with an
infrared divergence. This contribution was first calculated by
Crewther et al \cite{Crewther:1979pi}, and has been studied
thoroughly in the literature (see Fig. \ref{pionloop}). Diagrams (a)
and (b) in Fig. \ref{pionloop} contain an infrared divergence which
is regularized by the mass of pion and an analytical part. The
constant part is canceled by diagrams (c) and (d). Diagrams (e) and
(f) cancel with each other \cite{Pich:1991fq}.
Therefore, up to terms of order $(m_\pi/m_n)$, the neutron EDM
generated by the charged pion loop can be written as
\cite{Pich:1991fq}
\begin{equation}
d^{\rm n}_{\pi^+}=-\frac{e\sqrt{2}}{16\pi^2
F_\pi}h_c(D+F)\ln\left(m_\pi^2/m_n^2\right),
\end{equation}
where $D+F=-g_A=-1.26$ is the CP-even pion-nucleon coupling (the
signs of $D$ and $F$ is different from that in Ref.
\cite{Bernard:1995dp} because we are using a different definition of
the chiral transformation of $U$), and $h_c$ is the CP-odd
pion-nucleon coupling defined in Eq. (\ref{91}). Note that, in Fig.
\ref{pionloop}, the contribution from the proton's anomalous
magnetic moment has not been included. To include this contribution,
we consider all these diagrams in Fig. \ref{amdm} where the neutral
pion loop is also present, and the result is Ref. \cite{He:1992db}.
\begin{equation}
d^{n}_{\pi_0+\kappa}=\frac{e}{16\pi^2}\frac{D+F}{F_\pi}\left(-\sqrt{2}h_c\kappa_p+h_n\kappa_n\right)F_n\left(\frac{m_\pi^2}{m_n^2}\right)
\end{equation}
where $\kappa_n$ and $\kappa_n$ are tree-level anomalous magnetic moments of
protons and neutrons, respectively, and
\begin{equation}
F_n(s)=\frac{3}{2}-s-\frac{3s-s^2}{2}\ln
s+\frac{s(5s-s^2)-4s}{2\sqrt{s-s^2/4}}\arctan\frac{\sqrt{s-s^2/4}}{s/2}.
\end{equation}
We can see that there is no chiral enhancement in $F_n(s)$.

\begin{table}[h]
\begin{tabular}{|c|c|c|c|c|c|c|c|c|c|c|}
\hline &\multicolumn{9}{c|}{ \scriptsize{nEDM from different contributions / ($10^{-3}e C_4$ GeV)}}\\
\cline{2-10}  & \scriptsize{contact} &\scriptsize{meson photo-}&\multicolumn{2}{c|}{\scriptsize{$\pi$-$N$ coupling direct}}&\multicolumn{2}{c|}{\scriptsize{$\pi$-$N$ coupling, $\langle\pi^0\rangle,\;\langle\eta\rangle$}}&\scriptsize{CP-odd mass}&\scriptsize{CP-odd mass}&\\
\cline{4-7}
\scriptsize{operators}&\scriptsize{term}&\scriptsize{production}&\scriptsize{Fig.
\ref{pionloop}}&\scriptsize{Fig. \ref{amdm}}&\scriptsize{Fig.
\ref{pionloop}}&\scriptsize{Fig.
\ref{amdm}}&\scriptsize{direct}&\scriptsize{$\langle\pi^0\rangle,\;\langle\eta\rangle$}&\scriptsize{total}\\
\hline \scriptsize{$\bar u i\gamma_5 u \bar d
d$}&$-37.6$&$52.9$&$-30.9$&$71.3$&$27.2$&$202.1$&$15.8$&$-47.6$&$253.2$\\
\hline \scriptsize{$\bar u u\bar d i\gamma_5
d$}&$-62.6$&$-52.3$&$-30.9$&$-37.3$&$-30.2$&$-202.3$&$-25.1$&$30.6$&$-410.2$\\
\hline \scriptsize{$\bar u i\gamma_5 u\bar s
s$}&$0$&$52.9$&$0$&$0$&$27.2$&$202.1$&$0$&$-47.6$&$234.6$\\
\hline \scriptsize{$\bar u u\bar s i\gamma_5 s$}&$0$&$-0.6$&$0$&$0$&$2.8$&$-0.7$&$0$&$17.1$&$18.5$\\
\hline \scriptsize{$\bar d i\gamma_5 d \bar s
s$}&$0$&$-52.3$&$0$&$0$&$-30.2$&$-202.3$&$0$&$30.6$&$-254.3$\\
\hline \scriptsize{$\bar d d\bar s i\gamma_5
s$}&$0$&$-0.6$&$0$&$0$&$2.8$&$-0.7$&$0$&$17.1$&$18.5$\\
\hline \scriptsize{$\bar u i\gamma_5 u \bar u
u$}&$0$&$44.1$&$-30.9$&$17.0$&$22.7$&$168.9$&$0$&$-39.7$&$182.1$\\
\hline \scriptsize{$\bar d i\gamma_5 d\bar d
d$}&$0$&$-43.6$&$-30.9$&$-20.0$&$-25.1$&$-168.5$&$-18.6$&$25.5$&$-281.2$\\
\hline \scriptsize{$\bar s i\gamma_5s \bar s
s$}&$0$&$-0.5$&$0$&$0$&$2.4$&$-0.6$&$0$&$14.2$&$15.5$\\
\hline
\scriptsize{$\bar ui\gamma_5\sigma^{\mu\nu}u\bar d \sigma_{\mu\nu} d$}&$77.5$&$0$&$13.9$&$-7.7$&$0$&$0$&$2.1$&$0$&$85.8$\\
\hline \scriptsize{$\bar ui\gamma_5\sigma^{\mu\nu}u\bar
s\sigma_{\mu\nu}s$}&$0$&$0$&$0$&$0$&$0$&$0$&$0$&$0$&$0$\\
\hline \scriptsize{$\bar di\gamma_5\sigma^{\mu\nu}d\bar
s\sigma_{\mu\nu}s$}&\;\;\;\;$0$\;\;\;\;&$0$&\;\;\;\;\;$0$\;\;\;\;\;&$0$&\;\;\;\;\;\;$0$\;\;\;\;\;\;&$0$&$0$&\;\;\;\;\;\;$0$\;\;\;\;\;\;&$0$\\
\hline

\end{tabular}\caption{nEDM from the P-odd and CP-odd four-quark operators composed of color-singlet
currents. Different contributions are shown. }\label{singletedm}
\end{table}

\begin{table}[h]
\begin{tabular}{|c|c|c|c|c|c|c|c|c|c|c|}
\hline &\multicolumn{9}{c|}{ \scriptsize{nEDM from different contributions / ($10^{-3}e C_4$ GeV)}}\\
\cline{2-10}  & \scriptsize{contact} &\scriptsize{meson photo-}&\multicolumn{2}{c|}{\scriptsize{$\pi$-$N$ coupling direct}}&\multicolumn{2}{c|}{\scriptsize{$\pi$-$N$ coupling, $\langle\pi^0\rangle,\;\langle\eta\rangle$}}&\scriptsize{CP-odd mass}&\scriptsize{CP-odd mass}&\\
\cline{4-7}
\scriptsize{operators}&\scriptsize{term}&\scriptsize{production}&\scriptsize{Fig.
\ref{pionloop}}&\scriptsize{Fig. \ref{amdm}}&\scriptsize{Fig.
\ref{pionloop}}&\scriptsize{Fig.
\ref{amdm}}&\scriptsize{direct}&\scriptsize{$\langle\pi^0\rangle,\langle\eta\rangle$}&\scriptsize{total}\\
\hline \scriptsize{$\bar u i\gamma_5 t^a u \bar d
t^a d$}&$3.8$&$0$&$12.0$&$-17.6$&$0$&$0$&$-2.34$&$0$&$-4.2$\\
\hline \scriptsize{$\bar u t^a  u\bar d i\gamma_5 t^a
d$}&$8.9$&$0$&$12.0$&$4.42$&$0$&$0$&$6.0$&$0$&$31.2$\\
\hline \scriptsize{$\bar u i\gamma_5 t^a  u\bar s t^a
s$}&$0$&$0$&$0$&$0$&$0$&$0$&$0$&$0$&$0$\\
\hline \scriptsize{$\bar u  t^a u\bar s i\gamma_5 t^a
s$}&$0$&$0$&$0$&$0$&$0$&$0$&$0$&$0$&$0$\\
\hline \scriptsize{$\bar d i\gamma_5 t^a  d \bar s t^a
s$}&$0$&$0$&$0$&$0$&$0$&$0$&$0$&$0$&$0$\\
\hline \scriptsize{$\bar d t^a  d\bar s i\gamma_5 t^a
s$}&$0$&$0$&$0$&$0$&$0$&$0$&$0$&$0$&$0$\\
\hline \scriptsize{$\bar u i\gamma_5 t^a  u \bar u t^a
u$}&$0$&$-11.8$&$12.0$&$-6.6$&$-6.1$&$-45.0$&$0$&$10.6$&$-46.8$\\
\hline \scriptsize{$\bar d i\gamma_5 t^a  d\bar d t^a
d$}&$0$&$11.6$&$12.0$&$7.8$&$6.7$&$44.8$&$7.2$&$-6.8$&$83.4$\\
\hline \scriptsize{$\bar s i\gamma_5 t^a s \bar s t^a
s$}&$0$&$0.1$&$0$&$0$&$-0.6$&$0.2$&$0$&$-3.8$&$-4.1$\\
\hline
\scriptsize{$\bar ui\gamma_5\sigma^{\mu\nu} t^a u\bar d \sigma_{\mu\nu} t^a  d$}&$-103.1$&$0$&$93.8$&$-51.6$&$0$&$0$&$14.1$&$0$&$-46.9$\\
\hline \scriptsize{$\bar ui\gamma_5\sigma^{\mu\nu} t^a u\bar
s\sigma_{\mu\nu} t^a s$}&$0$&$0$&$0$&$0$&$0$&$0$&$0$&$0$&$0$\\
\hline \scriptsize{$\bar di\gamma_5\sigma^{\mu\nu} t^a d\bar
s\sigma_{\mu\nu}
t^a s$}&\;\;\;\;$0$\;\;\;\;&$0$&\;\;\;\;$0$\;\;\;\;&$0$&\;\;\;\;\;$0$\;\;\;\;\;&$0$&$0$&\;\;\;\;\;$0$\;\;\;\;\;&$0$\\
\hline
\end{tabular}\caption{Neutron EDM generated by P-odd and CP-odd four-quark operators composed of
color-octet currents. The labels have the same meaning as in Table \ref{singletedm}. }\label{octetedm}
\end{table}

Using the above, we estimate the pion-loop and the CP-odd
mass contributions to neutron EDM due to the P-odd and CP-odd four-quark operators.
The results are listed in Tables \ref{singletedm} and
\ref{octetedm}. Although the charged pion-loop (Fig. \ref{pionloop}) dominates in the chiral limit,
its numerical value is actually about an order of magnitude smaller
than the analytical chiral-loop contribution (Fig. \ref{amdm}). This is due to
the enhancement of $h_n$ relative to $h_c$ in the large $N_c$ limit.

The P-odd and CP-odd four-quark operators can also lead to
nonvanishing P-odd and CP-odd interaction like $n\rightarrow
K\Sigma$ and $n\rightarrow\Lambda\eta$. These interactions can
generate nEDM through kaon- or eta-loop diagrams. However, there is
no reason to believe that the kaon- or eta-loop contribution should
be more important than the pion-loop contribution so that it would
not change the order-of-magnitude estimate of nEDM generated by
those four-quark operators without the strange quark. For those
operators containing strange quark the estimation may not be
reliable and the kaon- or eta-loop contributions should be included.

\subsection{Comparison with Other Calculations and the Error-bars of this Calculation}

The P-odd and CP-odd four-quark contributions to neutron EDM have
been studied using different approximation methods in the literature
\cite{Khatsimovsky:1987fr,Valencia:1989cm,He:1992db}. The problem is
that it is difficult to get an estimate on the errors in any of
these methods. This is the strong motivation for the alternative
study presented here. By using a completely different approach, we
hope to get a better idea how well one actually estimates these
hadronic matrix elements.



In Ref. \cite{Khatsimovsky:1987fr}, the authors used the external
field method, factorization and QCD sum rules to make a direct
calculation of the neutron EDM. Their result is supposed to be the
total contribution, although it is unclear how the chiral physics
would be included in this approach. Their numbers are listed in
Table \ref{comparison} as ``factorization and QCD sum rule."  The
result is, in general, comparable to the charged pion-loop
contribution, although the contribution to the tensor operator is
particularly large.

In Ref. \cite{Valencia:1989cm}, the authors also calculated the
contributions of the pion-loop as we do in this paper. They used
entirely the factorization method to calculate the CP-odd
pion-nucleon couplings,  including the effects that the CP-odd
operators can annihilate the neutral pion in the vacuum. Taking the
operator $\bar u i\gamma_5 u\bar d d$ as an example, their
factorization works like this:
\begin{eqnarray}\label{facto}
\langle n\pi^0|\bar ui\gamma_5 u\bar d d|n\rangle&=&\langle n|\bar d
d|n\rangle\langle\pi^0|\bar u i\gamma_5 u|0\rangle\nonumber\\
&+&\langle 0|\bar d d|0\rangle\left(\langle n\pi^0|\bar u i\gamma_5
u|n\rangle - \frac{1}{m_\pi^2}\langle n\pi^0|{\cal
L}_{QCD}^m|n\pi^0\rangle\langle\pi^0|\bar ui\gamma_5u|0\rangle\right),
\end{eqnarray}
where ${\cal L}_{QCD}$ is the usual QCD Lagrangian. The terms inside
the bracket on the second line of the above formula cancel each
other. The reason is that $\bar ui\gamma_5 u$ is just a CP-odd mass
of the up-quark which can be rotated away through chiral
transformation, except for a possible $U_A(1)$ contribution. Thus
these two contributions should cancel with each other exactly. This
is first noticed in Ref. \cite{Donoghue:1986nv} in the spirit of the
Feinberg-Weinberg-Kabir theorem~\cite{Feinberg:1959ui}. Using this
method, one can get the CP-odd vertices, $h_c$ and $h_n$ as shown in
Table \ref{vevinducedh}. For the charged coupling $h_c$, one needs
to do a Fierz transformation, from which one can get a suppression
factor of $1/12$, where $1/3$ is from the color factor and the other
$1/4$ is from the spin. Therefore, $h_c$ is one order of magnitude
smaller than $h_n$. The corresponding nEDM calculated using this
method is included in Table \ref{comparison} as well.

\begin{table}[h]
\begin{tabular}{|c|c|c|c|}

\hline Operators & Our results & Naive factorization & Factorization and QCD sum rules\\
\hline $\bar u i\gamma_5 u \bar d d$&$253$&$-248$&$17.5$\\
\hline $\bar d i\gamma_5 d \bar u u$&$-410$&$177$&$-17.5$\\
\hline $\bar u i\gamma_5 d \bar s s$&$235$&$-85.8$&$-$\\
\hline $\bar s i\gamma_5 s \bar u u$&$18.5$&$0$&$-$\\
\hline $\bar d i\gamma_5 d \bar s s$&$-254$&$85.8$&$-$\\
\hline $\bar s i\gamma_5 s \bar d d$&$18.5$&$0$&$-$\\
\hline $\bar u i\gamma_5 u \bar u u$&$182$&$-154$&$-17.7$\\
\hline $\bar d i\gamma_5 d \bar d d$&$-281$&$203$&$15.2$\\
\hline $\bar s i\gamma_5 s \bar s s$&$15.5$&$0$&$-$\\
\hline $\bar u i\gamma_5\sigma^{\mu\nu} u \bar d\sigma_{\mu\nu} d$&$85.8$&$-79.4$&$-127.5$\\
\hline $\bar u i\gamma_5\sigma^{\mu\nu} u \bar s\sigma_{\mu\nu} s$&$0$&$0$&$-$\\
\hline $\bar d i\gamma_5\sigma^{\mu\nu} d \bar s\sigma_{\mu\nu} s$&$0$&$0$&$-$\\
\hline
\hline $\bar u t^a i\gamma_5 u \bar d t^a d$&$-4.2$&$-8.88$&$-3.18$\\
\hline $\bar d t^a i\gamma_5 d \bar u t^a u$&$31.3$&$-8.88$&$3.18$\\
\hline $\bar u t^a i\gamma_5 u \bar s t^a s$&$0$&$0$&$-$\\
\hline $\bar s t^a i\gamma_5 s \bar u t^a u$&$0$&$0$&$-$\\
\hline $\bar d t^a i\gamma_5 d \bar s t^a s$&$0$&$0$&$-$\\
\hline $\bar s t^a i\gamma_5 s \bar d t^a d$&$0$&$0$&$-$\\
\hline $\bar u t^a i\gamma_5 u \bar u t^a u$&$-46.8$&$39.5$&$-23.5$\\
\hline $\bar d t^a i\gamma_5 d \bar d t^a d$&$83.4$&$-51.1$&$9.3$\\
\hline $\bar s t^a i\gamma_5 s \bar s t^a s$&$ -4.12$&$0$&$-$\\
\hline $\bar u t^a i\gamma_5\sigma^{\mu\nu} u \bar d t^a \sigma_{\mu\nu}d$&$-46.9$&$-106$&$14.3$\\
\hline $\bar u t^a i\gamma_5\sigma^{\mu\nu} u \bar s t^a \sigma_{\mu\nu}s$&$0$&$0$&$-$\\
\hline $\bar d t^a i\gamma_5\sigma^{\mu\nu} d \bar s t^a \sigma_{\mu\nu}s$&$0$&$0$&$-$\\
\hline
\end{tabular}\caption{Comparison of different methods, nEDM calculated by factorization in
Ref. \cite{Valencia:1989cm,He:1992db} are shown as ``naive
factorization''. The column on the right side shows nEDM calculated
using factorization and QCD sum rules \cite{Khatsimovsky:1987fr}.
The unit of the numbers is $10^{-3} eC_4$GeV.}\label{comparison}
\end{table}


From Table \ref{comparison}, taking the operator $\bar
ui\gamma_5u\bar d d$ as an example, one can see that the magnitude
of our result is comparable with what obtained using na\"{\i}ve
factorization method but with a different sign; also the our result
is about one order of magnitude larger than the result estimated
using QCD sum rules. In our calculation, we separate the
contribution into the meson condensate contribution and the direct
matching contribution. The vacuum saturation method is used to
calculate the meson condensate contribution to $h_c$ and $h_n$. This
vacuum saturation method using to calculate the meson matrix
elements is accurate in the large-$N_C$ limit, which means the
calculation for this contribution is accurate up to
$1/N_C$~\cite{Manohar:1998xv}. From Table \ref{singletedm}, one can
see that the meson condensate contributions dominate over the direct
matching contributions. Therefore, for operators generating
unsuppressed meson condensates (see Sec IV for detailed
discussions), a conservative uncertainty can be set to be a factor
of two.

In Ref.~\cite{Valencia:1989cm}, the authors also used the vacuum
saturation approach to get the factorization result as shown in
Eq.~(\ref{facto}). However, in the case of baryon matrix element,
the non-factorized contribution is not suppressed in the large-$N_C$
limit~\cite{Manohar:1998xv}, therefore the missed non-factorized
contribution should be of the same order as the factorized
contribution shown in Eq.~(\ref{facto}). The calculation using QCD
sum rules in Ref.~\cite{Khatsimovsky:1987fr} did not include the
meson condensate contribution, therefore, their calculation might
miss an important contribution.

The factor of two uncertainty can also be seen from the
Feinberg-Weinberg-Kabir theorem~\cite{Feinberg:1959ui}. Applying to
this context, the theorem dictates that CP-odd $(3,\bar 3)$
two-quark operators give no contribution to CP-odd processes.
However, since we are using a hybrid method, this theorem may not be
satisfied. Therefore, the amount of violation of this theorem can be
seen as an estimate of the error of this calculation. Take the
operator $\bar ui\gamma_5u - \bar d i\gamma_5 d$ as an example,
following the prescription in Secs. III and IV, one can get
meson-condensate contribution to the neutral CP-odd pion-nucleon
coupling which can be written as
\begin{equation}
h_n^{mc} = \frac{2C_3(2c_1+c_3)}{F_\pi}\approx -\frac{10C_3}{F_\pi}\
,
\end{equation}
where $m_u = m_d = \bar m$ is assumed for the sake of simplicity,
$C_3$ is the Wilson coefficient of the two-quark operator and the
definitions of $c_1$ and $c_3$ can be found in Eq. (\ref{cs}). If
the $\sigma$-term is also employed to do the direct matching, one
can easily show that the direct matching contribution cancels the
meson condensate contribution exactly. Instead, in order to get the
uncertainty of our calculation we need to do the direct matching
using the quark model. Since the operator includes only products of
two quark fields, the calculation using the quark model is
straightforward, which gives
\begin{equation}
h_n^{dir} =  R \frac{3C_3}{F_\pi} \approx \frac{5C_3}{F_\pi}\ ,
\end{equation}
where the factor of 3 is due to that in the quark model the nucleon
contains three constituent quarks. $R\approx1.7$ comes from the
perturbative QCD effect as discussed in Sec. V. The anomalous
dimensions of the operator discussing here is as the same as the
anomalous dimensions of the quark mass. The relative sign between
the direct contribution and the meson condensate contribution is as
desired. However, the magnitude of the direct contribution is about
two times smaller than the meson condensate contribution. The
mismatch between the two contributions is due to that quark model
does not differentiate $\langle N|\bar q q|N\rangle$ and $\langle
N|q^\dagger q|N\rangle$. From this mismatch one can see that the
inaccuracy of the direct contribution calculated using quark model
might be a factor of two. Therefore, conservatively, the total
inaccuracy for those operators having unsuppressed vacuum condensate
contributions can be seen as a factor of two.

\begin{table}[h]
\begin{tabular}{lc}
\hline
Operators~&~Upper bound of $|C_4|/({\rm GeV}^{-2})$\\
\hline $\bar u i\gamma_5 u\bar d d$~&~$5\times10^{-12}$\\
$\bar u u\bar d i\gamma_5 d$~&~$4\times10^{-12}$\\
$\bar u i\gamma_5 u \bar s s$~&~$6\times10^{-12}$\\
$\bar d i\gamma_5 d \bar s s$~&~$6\times10^{-12}$\\
$\bar u i\gamma_5 u \bar u u$~&~$8\times10^{-12}$\\
$\bar d i\gamma_5 d \bar d d$~&~$5\times10^{-12}$\\
$\bar u i\gamma_5\sigma^{\mu\nu}u\bar d \sigma_{\mu\nu}
d$~&~$2\times10^{-11}$\\
$\bar u i\gamma_5 t^a u \bar d t^a d$ ~&~ $4\times10^{-10}$\\
$\bar u t^a u \bar d i\gamma_5 t^a d$ ~&~ $4\times10^{-11}$\\
$\bar u i\gamma_5 t^a u \bar u t^a u$ ~&~ $3\times10^{-11}$\\
$\bar d i\gamma_5 t^a d \bar d t^a d$ ~&~ $2\times10^{-11}$\\
$\bar ui\gamma_5\sigma^{\mu\nu}t^a u\bar d \sigma_{\mu\nu}t^a
d$~&~$3\times10^{-11}$\\
\hline
\end{tabular}\caption{Upper bound on the Wilson coefficients of P-odd, CP-odd four-quark operators, calculated using
the experimental data and hadronic matrix elements in this work.}\label{upper}
\end{table}

\section{CONCLUSION}

In this paper, we studied the four-quark contributions to the
neutron EDM, which dominate over other QCD operators in some new
physics models. Our approach was based on chiral expansion and
simple quark models. It is well known in the literature that the
leading chiral contribution comes from one-pion loop which dominates
in the chiral limit $m_\pi \rightarrow 0$, just like in the case of
the nucleon electric polarizability. Therefore, one needs to
calculate the four-quark contribution to the CP-odd pion nucleon
couplings. We studied these couplings in simple quark models, as an
alternate to large-$N_c$ factorization. We also considered ${\cal
O}(1)$ contribution from direct matching and pion-condensation to
the dipole moment, as well as the CP-odd nucleon mass contribution
through the magnetic moment. The resulting nEDM can be compared with
those from the naive factorization and QCD sum rules. The comparison
provides us some idea on the hadronic physics uncertainty in the
neutron EDM calculation. Our approach also provides a formalism for
lattice QCD calculations of the nucleon matrix elements of the
four-quark operators.

Using the matrix elements thus obtained, we obtain
new-physics-independent upper bounds on the Wilson coefficients of
four-quark operators from the experimental data. The current
experimental upper bound on neutron EDM is $2.9\times10^{-26}e$
cm~\cite{Baker:2006ts}. If we assume that there is no significant
cancelations among the contributions from these operators, we can
use the experiment limit to give upper bounds to the Wilson
coefficients of individual operators. In our calculation, the
strange quark effects were ignored, and we considered only operators
composed of up and down quarks. The final results are shown in Table
\ref{upper}.

It is interesting to note that the chiral-enhanced contribution
is actually large-$N_c$ suppressed. In fact, the non-singular
part of the chiral-loop contribution numerically dominates over
the singular one. This suggests a large-$N_c$ analysis
of the neutron EDM, including the delta resonance contribution.
However, this is beyond the scope of this paper.

\acknowledgments

We thank T. Cohen, P. Chen and Y. Zhang for useful discussions. This
work was partially supported by the U.S. Department of Energy via
Grant No. DE-FG02-93ER-40762. F. Xu acknowledges a scholarship
support from China's Ministry of Education.


\end{document}